\documentclass[pra, twocolumn, nofootinbib]{revtex4-2}
\pdfoutput=1
\usepackage[utf8]{inputenc}
\usepackage{hyperref}
\usepackage[british]{babel}
\usepackage{xcolor}
\usepackage{changes}
\usepackage{ragged2e} 
\usepackage{subcaption}
\DeclareCaptionJustification{justified}{\justifying}

\usepackage{graphicx}
\usepackage{hhline}
\usepackage{amsmath}
\usepackage{amssymb}
\usepackage{mathtools}
\DeclarePairedDelimiter{\ceil}{\lceil}{\rceil}
\DeclarePairedDelimiter{\floor}{\lfloor}{\rfloor}
\usepackage{braket}
\usepackage{xfrac}
\usepackage{makecell}

\begin{document}

\title{Minimizing resource overhead in fusion-based quantum computation\\ using hybrid spin-photon devices}

\author{Stephen C. Wein\textsuperscript{1}}
\author{Timothée Goubault de Brugière\textsuperscript{1}}
\author{Luka Music\textsuperscript{1}}
\author{Pascale Senellart\textsuperscript{2}}
\author{Boris Bourdoncle\textsuperscript{1}}
\author{Shane Mansfield\textsuperscript{1}}

\affiliation{\textsuperscript{1}Quandela, 7 Rue Léonard de Vinci, 91300 Massy, France}
\affiliation{\textsuperscript{2}Centre for Nanosciences and Nanotechnologies, CNRS, Université Paris-Saclay, UMR 9001, 10 Boulevard Thomas Gobert, 91120, Palaiseau, France}

\begin{abstract}
We present three schemes for constructing a (2,2)-Shor-encoded 6-ring photonic resource state for fusion-based quantum computing, each relying on a different type of photon source. We benchmark these architectures by analyzing their ability to achieve the loss tolerance threshold for fusion-based quantum computation using the target resource state. More precisely, we estimate their minimum hardware requirements for fault-tolerant quantum computation in terms of the number of photon sources to achieve on-demand generation of resource states with a desired generation period. Notably, we find that a group of 12 deterministic single-photon sources containing a single matter qubit degree of freedom can produce the target resource state near-deterministically by exploiting entangling gates that are repeated until success. The approach is fully modular, eliminates the need for lossy large-scale multiplexing, and reduces the overhead for resource-state generation by several orders of magnitude compared to architectures using heralded single-photon sources and probabilistic linear-optical entangling gates. Our work shows that the use of deterministic single-photon sources embedding a qubit substantially shortens the path toward fault-tolerant photonic quantum computation.
\end{abstract}
\maketitle

\section{Introduction}

Fault-tolerant quantum computation (FTQC) is vital to the correct execution of the most impactful quantum algorithms, such as prime factorization \cite{shor1994algorithms}, linear system solving \cite{harrow2009quantum}, and chemical simulations \cite{mcardle2020quantum}. Photonic architectures show particular promise for achieving fault-tolerance because photons are excellent carriers of quantum information. This property enables modularity, where numerous identical, smaller devices can be fabricated independently and interconnected to scale computational power. Such modularity is invaluable for improving the yield of components that meet the high quality standards necessary to satisfy the error correction thresholds \cite{aharonov1997fault}. Despite these advantages, the resource overheads currently required to solve industry-relevant problems remain a significant challenge \cite{kim2022fault}.

In photonic quantum computation, photon loss is the dominant error. For this reason, nearly all photonic fault-tolerant architectures focus on achieving high loss thresholds \cite{bartolucci2023fusion, paesani2023high, de2024spin, hilaire2024enhanced, chan2024tailoring}. One such architecture for FTQC is fusion-based quantum computation (FBQC) \cite{bartolucci2023fusion}. In this computing paradigm, which is a variation of measurement-based quantum computation \cite{briegel2009measurement}, photonic resource states are constructed and then measured. Notably, FBQC uses finite-sized graph states that are supplied to a fusion network, where all photons are consumed by two-qubit joint measurements, or fusion gates, that serve to both construct the photonic lattice and perform the desired quantum computation on it.

FBQC is fault-tolerant provided that the supplied resource states can be produced with a per-photon loss probability well below the loss-tolerance threshold, which depends on the fusion network, the choice of resource state, and the type of fusion gates. This threshold can be increased by code concatenation, where the qubits of the resource state are themselves encoded with an inner quantum error correcting code, such as a Shor code~\cite{bartolucci2023fusion, pankovich2024high, song2024encoded} or a graph code~\cite{bell2023optimizing}. This results in a bigger and more complex resource state. However, more complex states require more resources to construct that may also incur additional losses, therefore offsetting some of the original gain \cite{li2015resource}. It is thus crucial to explore different resource-state generator (RSG) architectures to identify strategies that minimize loss while also minimizing the resource overhead.

\begin{figure}
    \centering
    \includegraphics[width=0.95\linewidth, trim=80 195 70 130, clip]{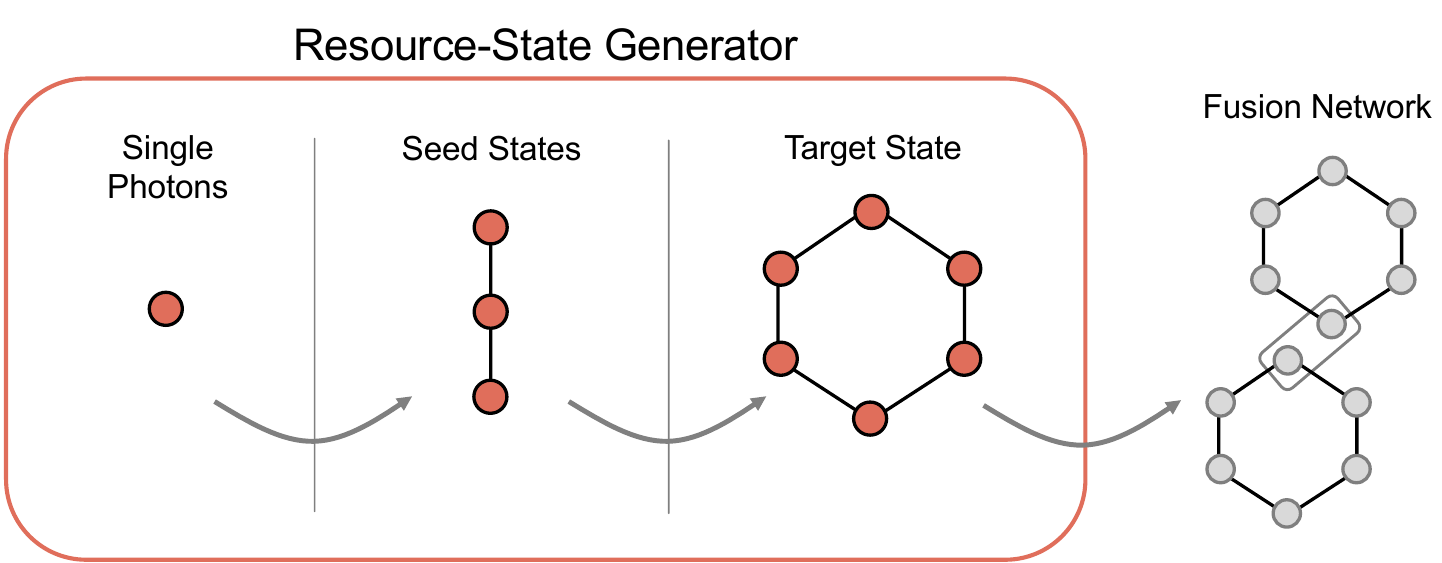}
    \caption{\textbf{Stages of fusion-based quantum computation.} Single photons can be combined using linear optics and measurements to produce entangled seed states, such as a linear-cluster state or a Greenberger–Horne–Zeilinger (GHZ) state. The black lines connecting photons indicate that a control-Z operation (CZ) was applied to the two photonic qubits each initially prepared in the $\ket{+}=(\ket{0}+\ket{1})/\sqrt{2}$ state. States which can be described using this representation are called graph states. Multiple seed states can be combined to construct a larger graph state, the target resource state, that is sent to a fusion network to implement fusion-based quantum computation. This work focuses on the stages up to and including the target state production.}
    \label{fig:rsg_general}
\end{figure}

In this work, we study the steps required to construct resource states for FBQC (see Figure \ref{fig:rsg_general}). We present three different architectures for constructing the 24-photon Shor-encoded (2,2) 6-ring resource state \cite{bartolucci2023fusion} that can achieve a loss tolerance of 7.5\% when used in conjunction with biased measurements \cite{bombin2303increasing}. The first architecture is all-photonic, requiring only a source of single photons, either heralded or deterministic. The second uses a deterministic source of caterpillar graph states \cite{huet2024deterministic}. The third uses a group of such sources that can be entangled using repeat-until-success (RUS) control-Z gates \cite{lim2005repeat}. This comparative analysis, which focuses on resource overhead and loss tolerance at the level of state generation, constitutes a critical first step to assess the main advantages provided by emitters for FBQC.

We begin in Section \ref{sec:sourcesarchi} by describing different source types and different schemes to process the photons they emit to construct the target resource state. In Section \ref{sec:comparison} we present the figures of merit we use to benchmark the architectures, compute their optimal bounds on resource efficiency and present simulation results. We end with a discussion in Section \ref{sec:discussion} and a brief outlook in Section \ref{sec:outlook}.

\section{Architectures for Photonic Resource-State Generation}
\label{sec:sourcesarchi}

We describe four types of single-photon sources and three schemes to manipulate the emitted photons. We call the combination of a source and a scheme an RSG architecture.

\begin{figure*}
    \centering
    \includegraphics[width=0.95\linewidth, trim= 30 100 20 75, clip]{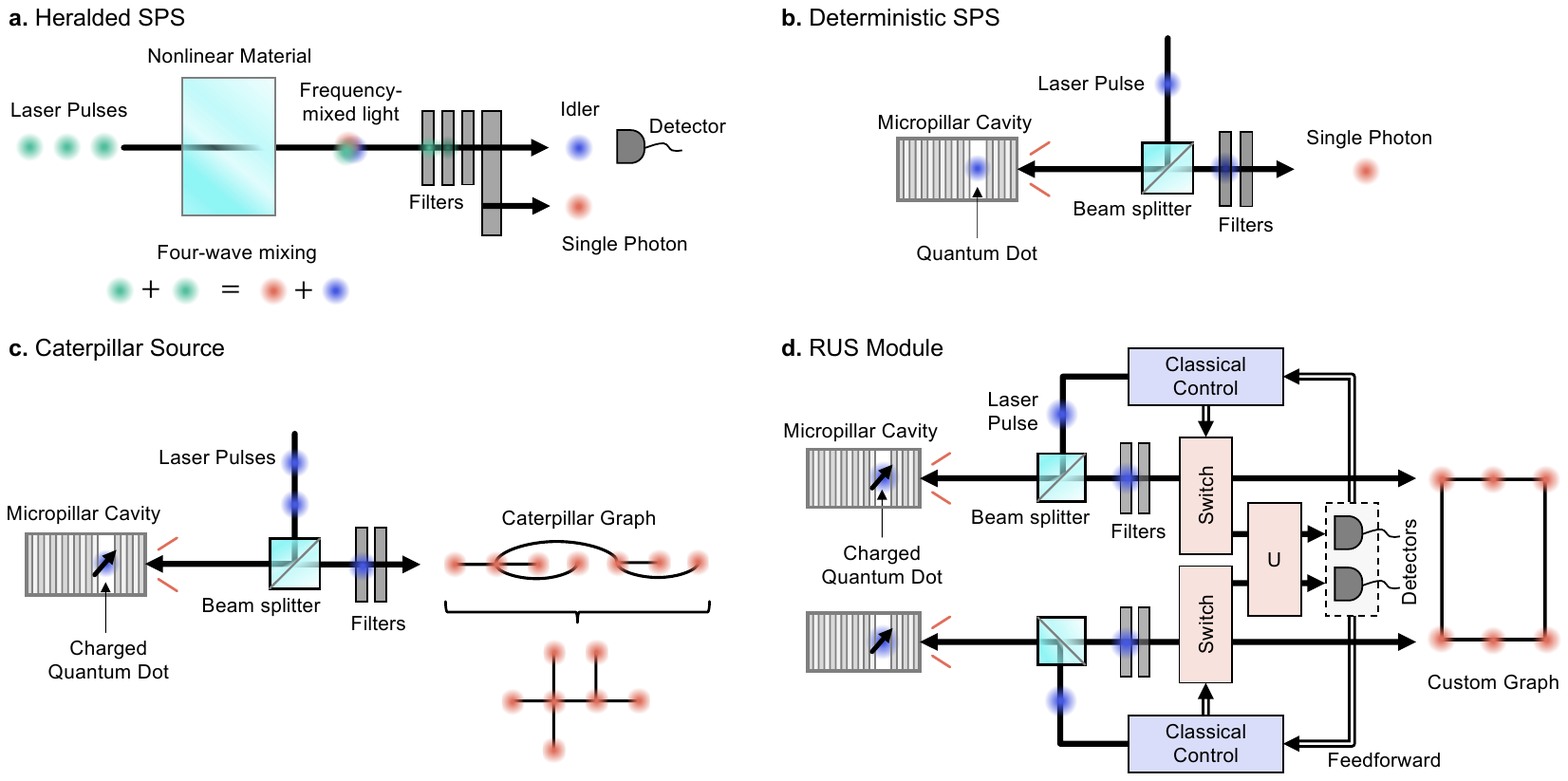}
    \caption{\textbf{Four different sources of quantum photonic states.} \textbf{a.} A heralded single-photon source (SPS) constructed using intense laser pulses interacting with a non-linear material to induce four-wave mixing, whereby two pump laser photons are converted into a pair of photons of different frequencies. By filtering the light in frequency and detecting the idler photon, the presence of the single-photon signal is heralded. \textbf{b.} A deterministic SPS constructed from a quantum dot (QD) embedded in a micropillar cavity. The QD is excited using off-resonant laser pulses and triggered to emit a single photon at its resonance frequency, which can be separated from scattered laser photons using a highly transmittive beam splitter and filter. \textbf{c.} A deterministic source of caterpillar graph states constructed from a deterministic SPS containing an additional matter qubit degree of freedom such as an electron spin. The source's qubit mediates entanglement between a chosen degree of freedom in successive photons such as polarization or time bin, producing a caterpillar graph state. \textbf{d.} A repeat-until-success (RUS) module where a switch is used to route photons entangled with the source's qubit to an entangling circuit implementing a unitary transformation $U$. Afterward, the photons are measured, and, upon success, a CZ gate is applied on the two source's qubits, thus allowing the construction of more complex graphs using fewer photons. In some case, local corrections on the source's qubits are required, which is handled by classical feedforward communication to the lasers controlling the sources.}
    \label{fig:sourcetypes}
\end{figure*}

\subsection{Photon sources}

\subsubsection{Heralded single-photon source}

A heralded single-photon source (HSPS) is a single-photon source (SPS) that succeeds in the generation of a single photon with probability $p_s<1$ when triggered, and provides a heralding signal when it does succeed. This can be achieved, for example, by exploiting four-wave mixing in a material with a third-order nonlinearity \cite{meyer2020single}. Upon excitation by a strong pump laser pulse, two laser photons in the pulse may spontaneously annihilate leading to the creation of a pair of photons with a probability $p_{\textrm{pair}}$, called the idler and the signal, with different frequencies that sum to the laser frequency (see Figure \ref{fig:sourcetypes}a). The idler and signal can then be split into two different spatial modes so that the detection of the idler heralds the presence of its twin photon in the other spatial mode. Taking into account the heralding efficiency $\eta_H$, which is the probability of detecting the idler given that a pair was produced, the probabilities $p_s$ and $p_{\textrm{pair}}$ are related by $p_s= \eta_H p_{\textrm{pair}}$ (for ease of reading, all the parameters we introduce and their associated symbols are summarized in Appendix~\ref{sec:recap-table}).

Increasing the pump power improves the success probability $p_s$, but, when the pump is too intense, it becomes very likely that unwanted multiple pairs are produced instead. This leads to a multi-photon error probability that is proportional to $p_s$. Thus, $p_s$ is commonly limited to below $1\%$. One technique proposed to overcome this issue relies on photon-number resolving detectors to measure the idler and discard multi-photon events~\cite{davis2022improved, sempere2022reducing}. In the limit of perfect heralding efficiency and photon-number resolution fidelity, this technique can boost the success probability up to at most $p_s=1/4$~\cite{Bonneau_2015}, provided that the heralding efficiency is sufficiently high (eg. $\eta_H>0.99$) to reach the desired noise suppression (see Appendix~\ref{appendix_g2}).

\subsubsection{Deterministic single-photon source}

A deterministic SPS (DSPS) does not suffer from a fundamental trade-off between emission probability and multi-photon errors. It produces a nearly perfect single photon when triggered and does not require a heralding signal. As such, it has a success probability of $p_s=1$, and any inefficiency is treated as photon loss and absorbed into the probability of collecting the photon from the source. Later on, we also include the photon collection efficiency as part of the total transmission efficiency (see \ref{subsubsec:trans_eff}).

Deterministic sources are commonly implemented using atoms \cite{kuhn2002deterministic} or artificial atoms such as quantum dots (QDs) \cite{senellart2017high}. The main challenge for this approach is designing and fabricating devices that efficiently collect the emitted photons. One very successful and robust approach is to fabricate microcavities using distributed Bragg reflectors to enhance QD emission into a single spatial mode \cite{somaschi2016near, ding2016demand}, that can then be collected by an optical fiber \cite{margaria2024efficient}. The QD emission can be triggered using excitation pulses that are detuned from the emission frequency of the QD \cite{thomas2021bright, bracht2021swing}, and thus can be separated from the emitted single photons using a filter (see Figure \ref{fig:sourcetypes}b). Alternatively, it is possible to excite the QD resonantly, collect emission using a birefringent cavity, and filter the orthogonally-polarized photons \cite{ding2023high}.

\subsubsection{Deterministic source of caterpillar graph states}

Deterministic emitters provide flexibility for generating useful states. Unlike heralded sources, the system's internal degrees of freedom can be exploited as a quantum memory to produce entangled states of light on demand through the process of sequential entanglement generation \cite{schon2005sequential, schon2007sequential, lindner2009proposal, thomas2022efficient}. The number of additional levels determine the complexity of the entangled state that can be produced \cite{schon2007sequential, thomas2024fusion}. By adding just one matter qubit degree of freedom, such as an electron spin, to a two-level emitter, it is possible to generate linear cluster states and GHZ states \cite{appel2022entangling,coste2023high, meng2024deterministic, su2024continuous}. More generally, such an emitter can produce photonic caterpillar graph states \cite{huet2024deterministic}, which are graph states where any dangling photons are at most one degree separated from a central one-dimensional cluster (see Figure \ref{fig:sourcetypes}c).

Caterpillar graph states are useful building blocks for constructing resource states. Since a caterpillar source can also produce GHZ states on demand, fusion gates can be boosted to a success probability of $1-1/2^b$, where the positive integer $b\geq 2$ is the level of boosting, while consuming $2^{b}-2$ additional auxiliary photons per gate \cite{grice2011arbitrarily}. Alternatively, $2b$ additional dangling photons can be added to the building block during generation to allow repeated fusion attempts without using auxiliary states, again bringing the success probability up to $1-1/2^b$ \cite{hilaire2023near}. This latter approach requires adaptivity, which can increase the chance that remaining dangling photons in the resource state are lost. In addition, to ensure a high success probability, many dangling photons must be produced even if the first fusion attempt ends up being successful. Since the gate will succeed after two attempts on average, many unused photons must be measured to decouple them from the resource state (See Appendix~\ref{sec:graph-prelim} for details on the fusion gates).

\subsubsection{Repeat-until-success module}
\label{subsubsec:RUSmodule}
To address the challenges of repeated fusions described above, an alternative is to entangle two or more caterpillar sources containing a qubit before emitting the fused graph (see Figure \ref{fig:sourcetypes}d for an example using two caterpillar sources). To entangle two sources, each source emits a photon that is sent to a small linear-optical module. When one of the correct detection patterns is observed, this results in a controlled-Z operation applied on the spin qubits in each source. The operation is thus probabilistic, but it can be repeated until it succeeds~\cite{lim2005repeat}. In the absence of photon loss, the probability of successfully producing an entangled state can then arbitrarily approach 1 at the cost of taking more time to attempt gates. 

This so-called RUS gate, and related RUS measurements, have recently been of great interest for designing hybrid spin-photon platforms for FTQC, such as the spin-optical quantum computing (SPOQC) architecture~\cite{de2024spin, hilaire2024enhanced} or the quantum-emitter-based FBQC scheme~\cite{chan2024tailoring}, showing impressive thresholds. Here, we demonstrate that it can also substantially improve a standard FBQC architecture, as the entanglement between the sources carries over to the emitted photons, thus allowing the generation of graph states with a more complex topology than caterpillar graphs, directly at source. Unlike repeated fusion gates, after entanglement is established, no additional dangling photons need to be produced. Therefore this approach ensures that a minimum number of photons are consumed to establish a graph edge. Moreover, the sources serve as a lossless quantum memory to store entanglement until other states are prepared. 

In our illustrations and discussions we will use charged quantum dots as a typical example to implement caterpillar sources and RUS modules. However, our study is also relevant for other emitters containing at least one qubit degree of freedom, such as T centers~\cite{simmons2024scalable}, SnV centers~\cite{chen2024scalable}, and atoms~\cite{thomas2024fusion}, to name a few.

\subsection{Three RSG schemes}
\label{subsec:architecture_desc}

\begin{figure*}
    \centering
    \includegraphics[width=0.90\linewidth, trim=70 80 40 70, clip]{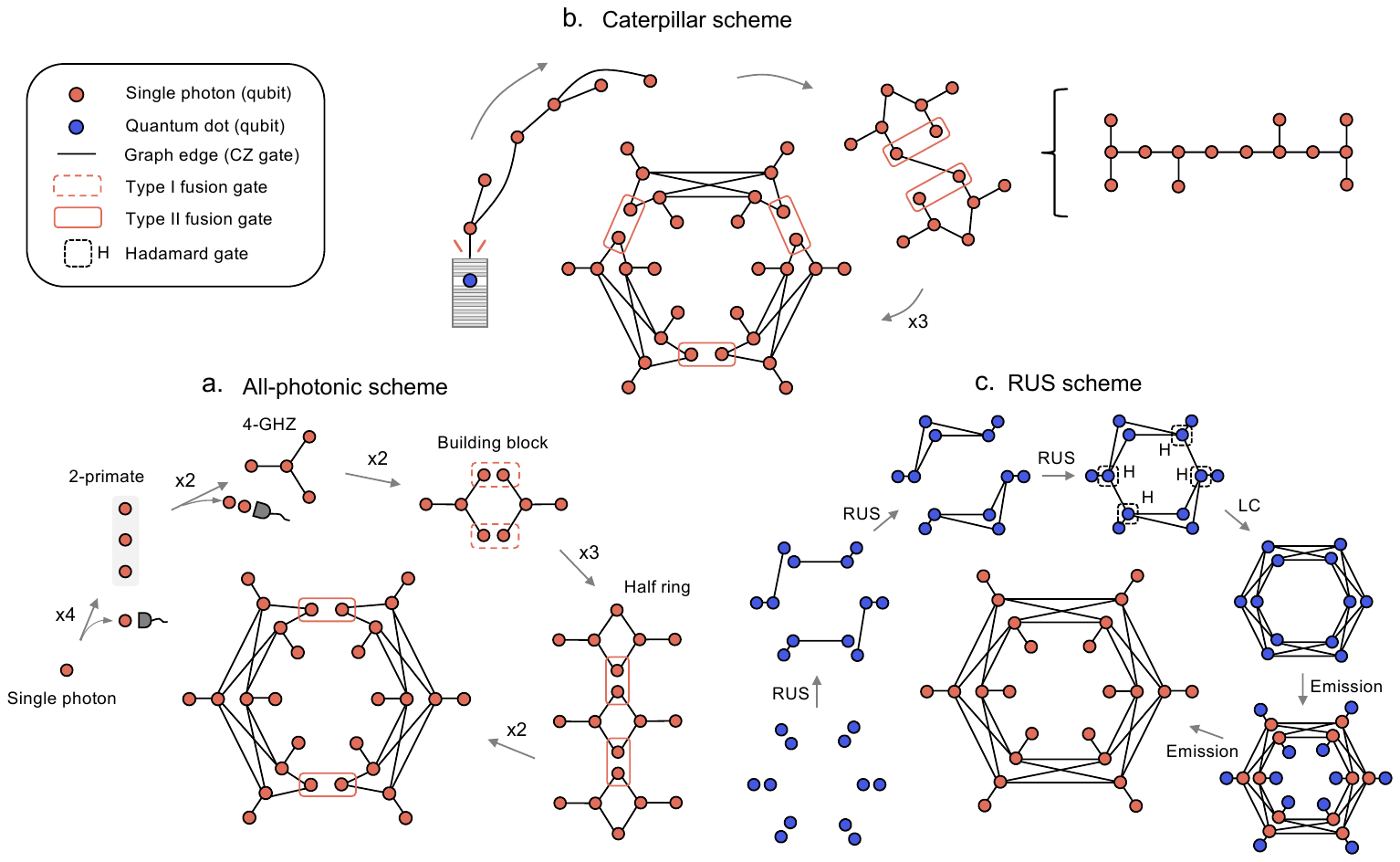}
    \caption{\textbf{Three different schemes for constructing a Shor-encoded (2,2) 6-ring resource state}. When several states need to be combined to progress to the next stage, the number is appended to the arrows. The fusion operations are boosted by generating and measuring additional auxiliary photons which are omitted for simplicity. \textbf{a.} An all-photonic architecture using single-photon sources that are either heralded or deterministic followed by probabilistic entanglement generation and fusion gates. \textbf{b.} Using a deterministic source of caterpillar graph states followed by fusion gates. \textbf{c.} Using a group of at least 12 deterministic caterpillar sources and repeat-until-success (RUS) CZ gates and local complementations (LC).}
    \label{fig:architectures}
\end{figure*}

An RSG scheme is a sequence of steps each involving multiple quantum operations which transform an initial state into the desired target resource state. The success or failure of each step is heralded by measurement outcomes. Linking the steps together involves spatial or temporal multiplexing. A complete RSG architecture is specified by combining an RSG scheme and a source producing the initial states which the scheme takes as input.

We now propose three different such RSG schemes to construct the 24-photon Shor-encoded (2,2) 6-ring resource state originally proposed in Ref. \cite{bartolucci2023fusion}. Each scheme aims to achieve as high a performance as possible by making the most of the capabilities of each source type presented earlier.

Each scheme also makes use of a tailored multiplexing strategy to overcome probabilistic operations. This technique achieves a high generation success probability by attempting probabilistic operations many times so that a sufficient number succeed. This repetition can be done in parallel using many sources distributed in space (spatial multiplexing) \cite{ma2011experimental} or using just a few sources and repeating operations sequentially in time (temporal multiplexing) \cite{kaneda2019high}. Spatial multiplexing costs more physical sources, while temporal multiplexing reduces the rate at which the RSG can produce resource states. Moreover, the successfully produced states must then be routed to the desired output mode(s) in both space and time, which incurs losses.

\subsubsection{All-photonic scheme}

The first approach is an all-photonic scheme that aligns with the original vision of FBQC, whereby multiple unentangled single photons are passed through probabilistic entangling linear-optical circuits and fusion gates that herald the successful generation of a resource state.

Due to probabilistic operations, as the resource state becomes more complex, the success probability exponentially drops to zero. A direct approach would then cause an unmanageable exponential resource overhead. This can be alleviated by dividing up the generation procedure into successive stages where intermediate operations are heralded \cite{li2015resource} so that complete states can be grouped together for the next operation. This dramatically increases the overall success probability and substantially decreases the amount of multiplexing needed to reach high-efficiency operation. The downside is that the number of lossy components seen by each photon, such as beam splitters and phase shifters, may increase to facilitate the selection and routing of photons between stages, possibly decreasing the maximal loss per component that the architecture can tolerate.

In Figure \ref{fig:architectures}a, we present an all-photonic scheme using six multiplexing stages. We designed this particular scheme to reduce the number of states needed to attempt a new stage while also maintaining a reasonable success probability for each stage. This provides a good compromise between boosting the success probability while minimizing the number of states being routed between stages. The construction begins with single photons, produced either probabilistically or deterministically. Four photons are combined to produce an intermediate 3-photon state, called a 2-primate \cite{bartolucci2021creation}. Two of these 2-primates can then be combined into a 4-photon GHZ states by measuring two of the photons. From there, twelve 4-GHZ states are fused together in three consecutive stages of multiplexing to produce the target resource state. Each resource produced during each of these final three stages requires two fusion gates to succeed simultaneously. We also consider the use of additional auxiliary photons to increase the success probability of these fusion gates (see Section~\ref{subsec:PerfBounds}).

This scheme can also be used starting with a source of 4-GHZ states. In that case, the first three stages of multiplexing can be neglected. We call the scheme that generates the 24-photon state directly from a source of 4-photon GHZ state a hybrid scheme.

\subsubsection{Caterpillar scheme}
\label{subsubsec:CaterpillarSource}

The caterpillar scheme first applies two type-II fusion gates to a 14-photon caterpillar graph to construct one-third of the resource state. The full 24-photon ring is then constructed by fusing three copies of those building blocks using three fusion gates (see Figure \ref{fig:architectures}b for the fusion operations and exact caterpillar state required).

Although only a single caterpillar source is needed to produce the initial 14-photon state, the following stages require fusion gates. Thus, in this architecture, while substantially reduced, multiplexing is still required to overcome the probabilistic nature of fusion operations. As before, to optimize the success probability, we consider boosted fusion gates, but this time using a Bell pair (not shown in Fig. \ref{fig:architectures}b), i.e. using Grice's scheme~\cite{grice2011arbitrarily}, because a caterpillar source can generate Bell pairs deterministically with no overhead.
 
\subsubsection{RUS scheme}
\label{subsubsec:rusmodulearchi}

The last approach that we introduce involves using RUS gates to entangle a small group of caterpillar sources. There are various ways to construct the resource state using interacting emitters \cite{russo2018photonic, russo2019generation, li2022photonic, takou2024optimization,barrett2005efficient}. Here we propose a scheme aimed primarily at minimizing the optical depth for the generated resource state. This is achieved by first creating all the necessary entanglement links between the sources and only emitting the photonic state once this first step is successful. In this way, the probability of losing a photon in the resource state is only impacted by the efficiency of the source and the transmission of a single active switch (Figure \ref{fig:sourcetypes}d). This active switch is necessary to switch between the RUS entangling stage and the emission stage. Losses due to the entangling optics for the RUS gate only impact generation efficiency rather than the final resource state.

We apply CZ gates between the emitters using the RUS module described in Section~\ref{subsubsec:RUSmodule}: each emitter sends a photon to a linear-optical interferometer, which implements a probabilistic rotated type-II fusion operation; if the fusion fails, we repeat the gate attempt, until we observe one of the correct detection patterns. Once the entanglement is built between the sources, it is directly transferred to the photons generated in the second stage (see Appendix~\ref{sec:graph-prelim} for details).

Naively generating all 32 edges of the resource state requires the detection of 128 photons on average. Losing just one photon can destroy many edges and substantially increase the generation time. This problem is alleviated by recognizing that the 12 dangling photons of the resource state can be produced deterministically, provided that the highly-connected central ring of 12 qubits is first prepared across 12 sources. Moreover, we exploit local complementation (LC)\footnote{LC is a graph operation that corresponds to complementing the neighborhood of a vertex in a graph, and when two graphs can be obtained from one another by LC, the corresponding graph states can be obtained from one another by local Clifford operations~\cite{PhysRevA.69.022316}.} to further reduce the number of edges that need to be established \cite{PhysRevA.83.042314}. Using the Graph State Compass (GSC) python package \cite{gsc}, we find that this 12-qubit graph has a minimum-edge representation with just 14 edges, which can be transformed into the desired 12-qubit central ring with four Hadamard gates, see Figure \ref{fig:architectures}c. The minimum-edge representation is hereafter called the LC-graph. This approach works only because the dangling photons can be produced afterwards, since the full 24-photon resource state is already in a minimum-edge representation.

The LC-graph has further useful properties. Most qubits have a low degree of connectivity and so photon loss is less destructive. In addition, it can be constructed using a divide-and-conquer strategy \cite{barrett2005efficient, lee2023graph} by connecting together identical subgraphs in three stages. Thus, if a photon is lost when generating one subgraph, it will not destroy the others.

Once the LC-graph is complete, it is transformed into the central 12-qubit ring of the resource state using LC. Each source then emits a photon, mapping the prepared graph onto the emitted photons (See Appendix~\ref{sec:graph-prelim} for details of this process). Since the sources remain entangled to the photonic graph, they can emit a second photon to produce the 12 dangling photons of the final resource state. Finally each spin is measured to decouple it from the graph and reset for the next cycle. 

\section{Architecture Efficiency Comparison}
\label{sec:comparison}

\subsection{Figures of Merit}

To compare architectures, we define two figures of merit quantifying the performance of an RSG: the resource efficiency $\eta_R$ and the transmission efficiency $\eta_T$. These metrics are designed to describe RSG architectures that operate on a periodic clock cycle with rate $r$, where a single resource state is produced by the RSG at the end of every cycle with an overall RSG success probability denoted as $p_\mathrm{rsg}$. This requirement is important to ensure synchronization between different RSGs and synchronization with the clock cycle of the fusion network required to perform FBQC.

\subsubsection{Resource efficiency}

To benchmark RSG overhead, a platform-agnostic metric is needed that captures trade-offs across RSG architectures. This metric could encompass factors like the energetic cost of operation, the fabrication cost, or the spatial footprint of the device. Designing an inclusive  and comprehensive definition of resource efficiency, as was initiated in~\cite{fellous2023optimizing}, is a challenge that should be addressed in a future work. Here, we consider a simple definition that enables the comparison of RSG architectures exploiting spatial and temporal multiplexing schemes.

To quantify the resource efficiency in the spatial domain, we use the ratio of the number of photons in the final resource state $N$ to the number of sources $N_0$ used to produce at least one such resource state with a desired probability $p_\mathrm{rsg}$. In the temporal domain, we consider the ratio of the RSG clock rate $r$ to the rate $r_0$ of the laser pump used to trigger the sources (the internal clock of the RSG). Combining these, we define a spatio-temporal resource efficiency of:
\begin{equation}
    \eta_R = \frac{Nr}{N_0r_0}.
    \label{eq:eta_r_2}
\end{equation}
By conservation of energy, we must always have $\eta_R\leq 1$. For example, one source ($N_0=1$) producing a large resource state of $N$ photons implies that $r \leq r_0/N$ as the source requires a minimum time to emit all $N$ photons. The ratio $\tau=r_0/r$ also quantifies the number of internal RSG clock cycles that the $N_0$ sources are operated for in order to attempt to produce a single resource state.

Crucially, the resource efficiency as defined above does not discriminate between spatial and temporal multiplexing. As such, it also does not capture important considerations regarding the range of values for $r$ and $r_0$ that are reasonable for a given material platform or fault-tolerant architecture. Thus, it is also important to interpret $\eta_R$ in the context of a specific implementation. We will touch on some of these considerations in Section \ref{sec:discussion}.

\subsubsection{Upper bound on $\eta_R$}
\label{sec:tradeoff}

The maximum possible resource efficiency, $\tilde{\eta}_R$, is fixed by the RSG architecture. This bound is reached in the limit of perfectly resource-efficient multiplexing \cite{bartolucci2021creation, bartolucci2021switch}. In this limit, an optimal multiplexer (MUX) is one that does not waste any photons. Every photon that is produced by a source will be consumed by the RSG to construct a resource state or released as part of the resource state\footnote{A non-optimal MUX may instead for example only be able to route one state if two are produced at the same time.}. Any excess photonic states produced during one RSG clock cycle may also be delayed to later cycles to help fill any deficit.

A spatial MUX does not decrease RSG rate ($r=r_0$) and so $\eta_R=N/N_0$ depends only on the number of sources $N_0$. If the architecture must consume an average of $N_\text{avg}$ photons to produce one resource state, then a necessary condition to arbitrarily improve the RSG success probability $p_\mathrm{rsg}$ toward $1$ is that the $N_0$ sources produce a minimum of $N_\text{avg}$ photons on average. Thus we have $p_sN_0\geq N_\text{avg}$ which implies $\eta_R\leq p_sN/N_\text{avg}=\tilde{\eta}_R$. Note that $p_s=1$ for all but the HSPS. This upper bound can be generalized to $r\neq r_0$ by noting that an optimal temporal MUX allows one to reduce $N_0$ at the cost of reducing $r$ by the same proportion. Hence:
\begin{equation}
    \eta_R=\frac{N r}{N_0 r_0} \leq \frac{p_s N}{N_\text{avg}}=\tilde{\eta}_R.
    \label{eq:eta_r}
\end{equation}

Knowing the upper bound, it is convenient to define the discrepancy between $\eta_R$ and $\tilde{\eta}_R$, which is directly attributed to the ability for a MUX to use the available photons. This leads to a natural definition of MUX efficiency $\eta_\text{mux} = N_\text{avg}/(p_sN_0)\cdot (r/r_0)$ so that
\begin{equation}
    \eta_R = p_s\eta_\text{mux}\frac{N}{N_\text{avg}}.
\end{equation}
This final expression is convenient as it clearly separates the three contributions to the resource efficiency: the success probability $p_s$ that depends only on the source, the MUX efficiency $\eta_\text{mux}$ that depends on the multiplexing network, and the efficiency of photon consumption $N/N_\text{avg}$ that depends only on the scheme. Note that $\eta_\mathrm{mux}$ is dependent on $p_s$ and $N_\mathrm{avg}$, since the optimal multiplexing scheme, and thus $\eta_\mathrm{mux}$, may change depending on the photon generation probability and overall scheme.

The above derivation of $\tilde{\eta}_R$ assumes that sources operate continuously for the entire RSG cycle. This is not the case for the RUS module because sources may idle after generating an edge. For example, in Figure \ref{fig:architectures}c, the last operation before LC will be a RUS gate to establish the edge connecting just two of the 12 sources. When this RUS gate is being applied, the 10 other sources must wait without emitting photons, which increases the number of internal cycles per source. Thus $N_\text{avg}$ will underestimate the cost in time and overestimate the upper bound on resource efficiency. In this case, the upper bound can instead be computed knowing the average number of internal clock cycles $\tau_\text{avg}$ used to produce a resource state: \begin{equation}
\label{eq:rusbound}    \eta_R\leq\tilde{\eta}_R=\frac{N}{N_0\tau_\text{avg}}.
\end{equation}
In the special case where each source attempts to produce a single photon at each internal cycle of the RSG, $N_\mathrm{avg}=p_sN_0\tau_\mathrm{avg}$ and we recover Eq. (\ref{eq:eta_r}).

\subsubsection{Transmission efficiency}
\label{subsubsec:trans_eff}

In practice, reaching the upper bound on resource efficiency is challenging, as improving $\eta_\text{mux}$ to unity often demands a complex multiplexing network capable of flexibly redistributing photons in space and time for use by the RSG at the right place and time \cite{bartolucci2021switch}. However, increasing the complexity of the multiplexing network reduces the transmission efficiency $\eta_T$ of photons in the RSG due to the greater number of optical components, such as beam splitters, phase shifters, mode crossings, and delay lines, required for photon routing. For this reason, it is useful to distinguish between the transmission efficiency that depends on the specific choice of multiplexing network, denoted here as $\eta_x$, and the transmission efficiency of photons produced by the source, which we denote $\beta$. Unlike $p_s$, which is non-unity only for HSPSs, $\beta$ impacts all types of sources and it quantifies the transmission between the source and the input of the RSG network, including the collection efficiency of the photons at the output of the source and loss due to filtering or frequency separation. This gives a total RSG transmission efficiency of
\begin{equation}
    \eta_T = \beta\eta_x.
\end{equation}

Note that both $\eta_T$ and the RSG success probability $p_\mathrm{rsg}$ impact the logical error rate of the fusion network but they do so in different ways. A non-unity $\eta_T$ will largely cause uncorrelated photon loss errors whereas non-unity $p_\mathrm{rsg}$ results from the failure of at least one probabilistic operation inside the RSG that may introduce correlated errors. The error-correction threshold for $\eta_T$ depends on the resource state being produced, and is known to be $92.5\%$ \cite{bombin2303increasing}. However, the error-correction threshold for $p_\mathrm{rsg}$ depends on how the state is constructed and would require a detailed error correction simulation of the fusion network to compute. Such a task goes beyond the scope of this comparative study and so we instead ensure $p_\mathrm{rsg} \geq 0.999\gg 0.925$ so that we explore a regime where errors are dominated by independent photon loss. In principle, FBQC with an appropriate decoding strategy could tolerate values of $p_\mathrm{rsg}$ lower than $0.999$, which could help increase $\eta_R$. However, as discussed in Appendix \ref{app:prsg}, the gain in $\eta_R$ is bounded: relaxing $p_\mathrm{rsg}$ from 0.999 to 0.9 increases $\eta_R$ by at most a factor of 3, and typically much less if $\eta_R$ is already close to $\tilde{\eta}_R$.

\subsubsection{Per-component maximal loss}

To achieve fault-tolerance, it is necessary to have $\eta_T\geq 0.925$. It is then instructive to find the minimum values of $\beta$ and the transmission of basic optical components contributing to $\eta_x$ that are needed to reach this target. 

Note that, if $\eta_x=1$, then we must have $\beta\geq 0.925$. Since the photon collection efficiency is characterized regularly in the literature \cite{ding2016demand, somaschi2016near, thomas2021bright, tomm2021bright, ding2023high, zhong201812, alexander2024manufacturable}, we do not break down the various components contributing to it. Instead, we define it as the probability of having a photon in a waveguide or fiber at the point just after any necessary filters (recall Figure \ref{fig:sourcetypes}).

Setting aside $\beta$ for now, we analyze $\eta_x\geq 0.925$. This requirement is satisfied by many possible configurations of the RSG that each correspond to a different value of $\eta_R$. It is thus important and nontrivial to find the maximum resource efficiency such that the RSG operates in the loss-tolerant regime. For a given architecture, and considering only $D$ optical components that dominate the losses with a similar loss-per component of $\sim x$, we can write $\eta_x=(1-x)^D$. Here, we denote $D$ as the optical depth of the network, or the number of the most-lossy components seen by a photon passing through the entire network to be released by the RSG as part of the final resource state. Importantly, we estimate in Section \ref{subsec:PerfBounds} below a lower bound on $D$ for each architecture, which will imply a maximal per-component loss $x$ such that $\eta_x\geq 0.925$ using:

\begin{equation}
\label{eq:threshold} 
    x\leq 1 - 0.925^{1/D}.
\end{equation}

\subsection{Performance bounds}
\label{subsec:PerfBounds}
\begin{table*}[ht]
\centering
\begin{tabular}{cc|cc|ccc}
\hline
\multicolumn{1}{c|}{Source} &
  \begin{tabular}[c]{@{}c@{}}Generation\\ scheme \end{tabular} &
  \multicolumn{1}{c|}{\begin{tabular}[c]{@{}c@{}}MUX stages\\ $N_\text{mux}$\end{tabular}} &
  \begin{tabular}[c]{@{}c@{}}Optical depth\\ $D$  \end{tabular} &
  \multicolumn{1}{c|}{\begin{tabular}[c]{@{}c@{}}Resource \\ efficiency $\tilde{\eta}_R$ \end{tabular}} &
  \multicolumn{1}{c|}{\begin{tabular}[c]{@{}c@{}}Number of SPS\\ $N_0$ \end{tabular}} &
  \begin{tabular}[c]{@{}c@{}}Maximal loss per \\ component $x$ \end{tabular} \\ \hhline{=|=|=|=|=|=|=}
\multicolumn{1}{c|}{HSPS, $p_s$=0.005} &
  All-photonic &
  \multicolumn{1}{c|}{$6$} &
  $36$ &
  \multicolumn{1}{c|}{$0.00057\%$} &
  \multicolumn{1}{c|}{$423400$} &
  $0.22\%$ \\
\multicolumn{1}{c|}{HSPS, $p_s$=0.05} &
  All-photonic &
  \multicolumn{1}{c|}{$6$} &
  $36$ &
  \multicolumn{1}{c|}{$0.0057\%$} &
  \multicolumn{1}{c|}{$42340$} &
  $0.22\%$ \\
\multicolumn{1}{c|}{HSPS, $p_s$=0.25} &
  All-photonic &
  \multicolumn{1}{c|}{$6$} &
  $36$ &
  \multicolumn{1}{c|}{$0.029\%$} &
  \multicolumn{1}{c|}{$8468$} &
  $0.22\%$ \\
\multicolumn{1}{c|}{DSPS} &
  All-photonic &
  \multicolumn{1}{c|}{$5$} &
  $30$ &
  \multicolumn{1}{c|}{$0.11\%$} &
  \multicolumn{1}{c|}{$2117$} &
  $0.26\%$ \\
\multicolumn{1}{c|}{4-GHZ source} &
  Hybrid &
  \multicolumn{1}{c|}{$3$} &
  $18$ &
  \multicolumn{1}{c|}{$0.77\%$} &
  \multicolumn{1}{c|}{$310$} &
  $0.43\%$ \\
\multicolumn{1}{c|}{Caterpillar source} &
  Caterpillar &
  \multicolumn{1}{c|}{$2$} &
  $12$ &
  \multicolumn{1}{c|}{$9.9\%$} &
  \multicolumn{1}{c|}{$25$} &
  $0.65\%$ \\
\multicolumn{1}{c|}{RUS Module} &
  RUS &
  \multicolumn{1}{c|}{$0$} &
  $1$ &
  \multicolumn{1}{c|}{$20\%$} &
  \multicolumn{1}{c|}{$12$} &
  $7.5\%$ \\ \hline
\end{tabular}%
\caption{\textbf{Performance bounds to generate a Shor-encoded (2,2) 6-ring resource state with the proposed architectures}. For each source and generation scheme, the third column gives the number of multiplexing stages and the fourth column gives the estimated minimum optical depth associated to the multiplexing network. The maximum resource efficiency in the fifth column corresponds to Eq.~\eqref{eq:eta_r}, except for the RUS module, for which the bound is given by Eq.~\eqref{eq:rusbound}, and an example of number of SPS reaching this bound is computed via Eq.~\eqref{eq:eta_r_2} for a rate $r=100$MHz and displayed in the sixth column. The last column gives the maximal loss per component that each architecture can tolerate to reach a total transmission efficiency of $92.5\%$~\cite{bombin2303increasing} when $\beta=1$, given by Eq.~\eqref{eq:threshold}. Since $\beta$ contributes only once to the total transmission, this maximal loss per component also estimates the required source efficiency to be $\beta\geq 1 - x$. The non-RUS architectures use fusion gates boosted to a success probability of $3/4$. All sources are taken to operate at a base repetition rate of $r_0 = 1$ GHz, which is expected to be limited by recovery times for detectors and feedforward operations.}
\label{tab:optimal}
\end{table*}

Using the above-defined metrics along with the assumption of perfectly efficient multiplexing ($\eta_\text{mux}=1$), we benchmark the bounds on performance for different sources combined with the schemes described in Section~\ref{subsec:architecture_desc}:
\begin{enumerate}
    \item
    \begin{enumerate}
        \item all-photonic scheme with a HSPS at $p_s=0.5\%$ representing current technology,
        \item all-photonic scheme with a HSPS at $p_s=5\%$ representing realistic improvements,
        \item all-photonic scheme with a HSPS at $p_s=25\%$ representing the most optimistic scenario,
        \item all-photonic scheme with a DSPS,
        \item a hybrid scheme that uses a deterministic source of 4-GHZ states to bypass the first steps in the all-photonic scheme,
    \end{enumerate}
    \item the caterpillar source and scheme described in Section~\ref{subsubsec:CaterpillarSource}, and
    \item the RUS module and scheme described in Section~\ref{subsubsec:rusmodulearchi}.
\end{enumerate}

The results are summarized in Table \ref{tab:optimal}.

\subsubsection{All-photonic scheme}

The upper bound on resource efficiency $\tilde{\eta}_R$ for the all-photonic scheme can be computed from Eq. \ref{eq:eta_r} by first computing the average number of photons $N_\text{avg}$ consumed to produce one resource state on average. This can be computed following the method in Ref. \cite{bartolucci2021creation} by knowing the success probability $p_i$ of each stage $i=1,2,...$ of the six-stage scheme, along with the number of auxiliary single photons $a_i$ used to boost fusion gates and the number of copies $c_i$ of the state required for the next stage.

For example, to produce on average one 2-primate we need $c_1=4$ single photons per attempt and we must take on average $1/p_2\simeq3.5$ attempts where $p_2=20\sqrt{2}-28\simeq 28.4\%$ \cite{bartolucci2021creation}. Hence, a 2-primate state requires on average $3.518\times 4 \simeq 14.072$ photons to produce. The 4-GHZ state then requires $c_2=2$ 2-primate states costing a total of 28 photons on average, and it succeeds with a probability of $p_3=(3+2\sqrt{2})/64\simeq 9.11\%$ corresponding to an average of $1/p_3\simeq 10.98$ attempts. Thus, the average number of photons needed to produce one 4-GHZ state on average is $\sim 309$, as originally derived in Ref. \cite{bartolucci2021creation}.

In Appendix~\ref{sec:appNAvg}, we extend this method to any number of MUX stages. This gives a recursive formula to compute the cumulative average number of photons consumed to succeed the $i$th stage:
\begin{equation}
\label{eq:avg}
N_{\text{avg},i} = c_i\frac{N_{\text{avg},i-1}+a_i}{p_i},
\end{equation}
where $N_{\text{avg},0}=1$. Similar recursive formulas have been derived in previous work to estimate resource overhead for photonic state construction \cite{lee2023graph}.

The values of $p_i$, $c_i$ and $a_i$ for the six-stage all-photonic scheme are summarized in Table \ref{tab:allphotonicsummary}. Note that standard fusion gates succeed with a probability of $50\%$. Here, to improve the probability of successfully generating the state, we assume that each gate is boosted to reach a success probability of $75\%$ using four additional single photons (not shown in Fig.~\ref{fig:architectures}a) \cite{bartolucci2021creation}. Indeed, in the low-loss regime considered in our study, the cost of producing additional photons is lower than the cost of multiplexing more fusion gates. However, boosting beyond $75\%$ requires additional GHZ states, which are themselves costly to produce from single photons and do not justify the mild increase to success probability compared to increasing the amount of multiplexing.

Following Eq. (\ref{eq:avg}) using the values in the table, we arrive at the total of $N_\text{avg}=21170$ single photons for one resource state. Finally, from Eq. (\ref{eq:eta_r_2}), we have the upper bound
\begin{equation}
\tilde{\eta}_R = \frac{24p_s}{21170}
\end{equation}

For a heralded SPS with $p_s=0.5\%$ success probability, the resource efficiency is then bounded by $\eta_R\leq 0.00057\%$, while for a deterministic source it is $\eta_R\leq 0.11\%$. Assuming that the internal clock cycle (the pump repetition rate) is $r_0=1$ GHz and we wish to produce resource states at the rate $r=100$ MHz, these resource efficiencies correspond to a minimum number of sources of $N_0 = 423400$ and $N_0 = 2117$, for the heralded and deterministic SPSs respectively.

\begingroup
\renewcommand{\arraystretch}{1.7}
\begin{table}
    \centering
    \begin{tabular}{c||c|c|c|c|c|c}
    \hline
         Stage, $i$& 1 & 2 & 3 & 4 & 5 & 6 \\\hline\hline
        Probability, $p_i$ & $1$ & $20\sqrt{2}-28$ & $\dfrac{3+2\sqrt{2}}{64}$ & $\dfrac{9}{16}$ & $\dfrac{9}{16}$ & $\dfrac{9}{16}$ \\
         Copies, $c_i$& 4 & 2 & 2 & 3 & 2 & 1\\
         Auxiliaries, $a_i$& 0 & 0 & 0 & 8 & 8 & 8 \\
         Detected, $d_i$ & 1 & 1 & 2 & 10 & 12 & 12 \\
         \hline
         Cumul. $N_{\text{avg},i}$ & $4$ & $28$ & $618$ & $3339$ & $11900$ & $21170$\\
         Cumulative\ $D_i$ & $9$ & $15$ & $21$ & $27$ & $33$ & $36$ \\ \hline
    \end{tabular}
    \caption{A summary of the relevant quantities describing the multiplexing stages of the proposed all-photonic scheme to building a Shor encoded (2,2) 6-ring resource state. Here, $p_i$ is the success probability of the $i$th stage operation in the absence of loss, $c_i$ is the number of copies needed of the state produced in that stage to attempt the following stage, $a_i$ is the number of auxiliary single photons needed to boost fusion gates using Ewert and van Loock's scheme~\cite{ewert20143}, $d_i$ is the number of detected photons, and \ $N_\text{avg, $i$}$ is the cumulative average number of single photons needed to succeed the stage. $D_i$ is the cumulative optical depth needed to construct the resource state and to route the photons through the network.}
    \label{tab:allphotonicsummary}
\end{table}
\endgroup

To estimate the maximal loss per component $x$ required to remain in the fault-tolerant regime, we first make an optimistic approximation that the optical depth of a photonic MUX does not depend on the number of photonic states being routed between each stage. That is, we assume that the optical depth of the RSG depends only on the type of MUX networks used and the number of MUX stages. For the type of spatial MUX network, we consider a Spanke network that uses two layers of generalized Mach-Zehnder interferometers (GMZIs) with a mode routing network in between \cite{bartolucci2021switch}. Each GMZI is composed of many beam splitters, but each photon only sees one phase shifter using the most optimistic design \cite{bartolucci2021switch}. Since a phase shifter is expected to be the most lossy component for a GMZI, we estimate the minimum optical depth to be $D_\text{GMZI}=1$. The mode routing network requires photons to be redistributed in space, which requires long propagation distances and for many lossy mode crossings. For this reason, the practical solution is to couple photons into low-loss fibers to redistribute them arbitrarily with minimal propagation loss. However, each chip-to-fiber interface causes non-negligible loss, which we identify as the most lossy component. Since it is necessary to couple out of the first GMZI and into the second GMZI, we estimate the minimum constant optical depth of the mode routing network to be $2D_\text{coupl.}=2$.

To route photons through the network, it is also necessary to reconfigure it in between each internal clock cycle. During this time, photons must be stored in a low-loss memory. Assuming that the cycle time is at least 1 ns to allow for classical processing, and knowing that state-of-the-art propagation loss in silicon nitride (SiN) is around 0.4 dB/m, the loss from a single on-chip memory is at least 60 mdB (or around 1.4\% loss). Given that chip-to-fiber couplings can already be less than 60 mdB \cite{alexander2024manufacturable}, a fiber memory is more practical and allows for delays much longer than 1 ns without significant loss. Thus, we also assume that every MUX stage begins and ends with a chip-to-fiber interface, bringing the minimum optical depth per spatial MUX stage to $D_\text{s}=2D_\text{GMZI}+4D_\text{coupl.}=6$.

To allow for temporal multiplexing, we consider the de Bruijn-based spatio-temporal network introduced in \cite{bartolucci2021switch}, which is similar to the Spanke network but includes an additional GMZI and a layer of delay lines to synchronize photons in time. In this case, the delay lines must store photons longer than 1 ns so that they can be redistributed over many internal clock cycles, leaving no choice but to implement them using fiber memories. Hence, we consider a spatio-temporal MUX to include an additional optical depth of $D_\text{GMZI}+2D_\text{coupl.}=3$ for a total of $D_\text{b}=9$.

Spatio-temporal multiplexing is a useful technique to reduce the number of sources while also allowing control over the RSG rate $r$ independent of the internal rate $r_0$. To gain this flexibility, it is only necessary to use it for one stage while the remaining stages only use a spatial MUX strategy. Furthermore, the final stage of multiplexing requires only a $1\times n$ MUX, which can be accomplished in depth 3. This brings us to our final lower-bound estimate on the optical depth of an RSG:
\begin{equation}
    D = D_{b} + (N_\text{mux} - 2)D_s + 3
    \label{formula_depth}
\end{equation}
where $N_\text{mux}>1$ is the number of multiplexing stages for the scheme. Thus for the full six-stage scheme, the minimum optical depth is $D=36$, leading to a per-component maximal loss of $x=1-0.925^{1/36} = 0.22\%$.

\subsubsection{Caterpillar scheme}

\begingroup
\renewcommand{\arraystretch}{1.7}
\begin{table}
    \centering
    \begin{tabular}{c||c|c|c}
    \hline
         Stage, $i$& 1 & 2 & 3\\
         \hline\hline
        Probability, $p_i$ & 1 & $\dfrac{9}{16}$ & $\dfrac{27}{64}$ \\
         Copies, $c_i$& 1 & 3 & 1 \\
         Auxiliaries, $a_i$& 0 & 4 & 6 \\
         Detected, $d_i$ & 0 &8 & 12 \\
         \hline
         Cumulative\ $N_{\text{avg},i}$ & 14 & 96 & 242 \\\
         Cumulative\ $D_i$ & 0 & 9 & 12 \\
         \hline
    \end{tabular}
    \caption{A summary of the relevant quantities describing the multiplexing stages of the proposed caterpillar scheme to building a Shor encoded (2,2) 6-ring resource state. Here, $p_i$ is the success probability of the $i$th stage operation in the absence of loss, $c_i$ is the number of copies needed of the state produced in that stage to attempt the following stage, $a_i$ is the number of auxiliary photons needed to boost fusion gates using Grice's scheme~\cite{grice2011arbitrarily}, $d_i$ is the number of detected photons, and $N_\text{avg,$i$}$ is the cumulative average number of single photons needed to succeed the stage. $D_i$ is the cumulative optical depth needed to construct the resource state and to route the photons through the network.}
    \label{tab:caterpillarsummary}
\end{table}
\endgroup

For the caterpillar scheme, we can compute the average number of photons $N_\text{avg}$ required to construct a single resource state using the same formula as for the all-photonic scheme. The probabilities, number of copies, and auxiliary photons used for fusion gates are all summarized in Table \ref{tab:caterpillarsummary}. Here, we assume that each fusion is again boosted to a 75\% success probability, but this time using an auxiliary Bell pair \cite{grice2011arbitrarily}. Note that a caterpillar source can also be used to efficiently boost fusion gates beyond 75\%, however we find that the relative improvement compared to once-boosted gates is minor (see Appendix~\ref{apx:bounds}) and may not justify the additional losses due to the increased complexity. 
Hence, we focus on the once-boosted fusion gate in our analysis.

Applying Eq. \ref{eq:avg}, we find that the average number of photons consumed to produce one resource state on average is $N_\text{avg}=242$. This corresponds to an upper bound on the resource efficiency of $\tilde{\eta}_R=9.9\%$ for the $N=24$-photon state.

For the minimum optical depth, we assume the architecture uses the same MUX networks as for the all-photonic architectures, but now using $N_\text{mux}=2$. This gives a minimum optical depth of $D=12$ for the caterpillar scheme, including one spatial-temporal MUX with depth 9 and one final spatial MUX with depth 3. This leads to a per-component maximal loss of $x=0.65\%$.

\subsubsection{RUS scheme}

To estimate the performance bounds for the RUS module architecture, we first assume that exactly $N_0=12$ sources are used so that the probabilistic nature of the CZ gates is only overcome using RUS, which is a form of temporal multiplexing, rather than photonic spatial multiplexing. Then, we count the average number of gate attempts made per source to estimate $\tau_\text{avg}$ for the full scheme, which then gives the upper bound using Eq. (\ref{eq:rusbound}).

Each gate attempt succeeds with probability $1/2$ \cite{lim2005repeat,de2024spin, hilaire2024enhanced}, therefore each round of RUS gates requires 2 attempts on average. Since the maximum degree of the LC-graph is 4 (recall Figure \ref{fig:architectures}c), 4 rounds of RUS gates are necessary to produce the spin graph. Thus, all 12 sources must wait for an average of 8 RUS attempts before the spin graph is complete. Finally, we account for two emission cycles that we assume operate at the same rate $r_0$ for simplicity. In total, the resource state can be generated using on average $\tau_\text{avg}=10$ internal clock cycles. Using Eq. (\ref{eq:rusbound}), this gives the RUS module approach a maximum resource efficiency of $\tilde{\eta}_R=24/(12\times 10) = 20\%$. Note that this bound neglects the time needed to implement local operations such as single-qubit gates and measurements. In practice, these operations can be implemented much faster than the feedforward-limited RUS gates, and so we assume they have a negligible impact on $\tau_\text{avg}$ (see Appendix~\ref{app:MonteCarlo} for more discussion on operation timescales).

As discussed in Section \ref{subsubsec:rusmodulearchi}, the RUS scheme was designed to minimize the optical depth to a single $1\times 2$ switch. In addition, although some platforms may use chip-to-fiber coupling to route photons through the switch, this is not strictly necessary because photons are stored in the spin memory until the switch is reconfigured. A single switch is also significantly more shallow than a multiplexing stage needed to route probabilistic inputs. For these reasons, we estimate a minimum optical depth of $D=1$ for the RUS scheme. This implies that the per-component maximal loss is $x=7.5\%$, which is significantly higher than all other architectures thanks to the lossless memory of the RUS module. Note that, since $7.5\%$ is also substantially higher than the anticipated loss from a single active switch, the loss will be dominated by the source itself, meaning that a source with $\beta\geq 92.5\%$ reaches the loss-tolerant regime for this architecture.

\subsection{Implementations}
\label{sec:concrete}

\begin{figure*}[t!]
    \centering
    \hspace{-50mm}\textbf{a.} \hspace{85mm} \textbf{b.} \\
    \includegraphics[width=0.47\linewidth, trim= 0 0 0 0, clip]{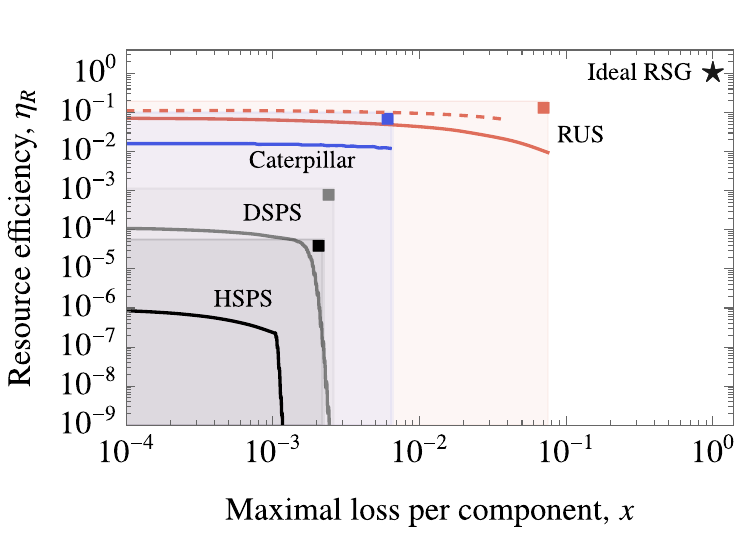} 
    \qquad
    \includegraphics[width=0.47\linewidth, trim= 0 0 0 0, clip]{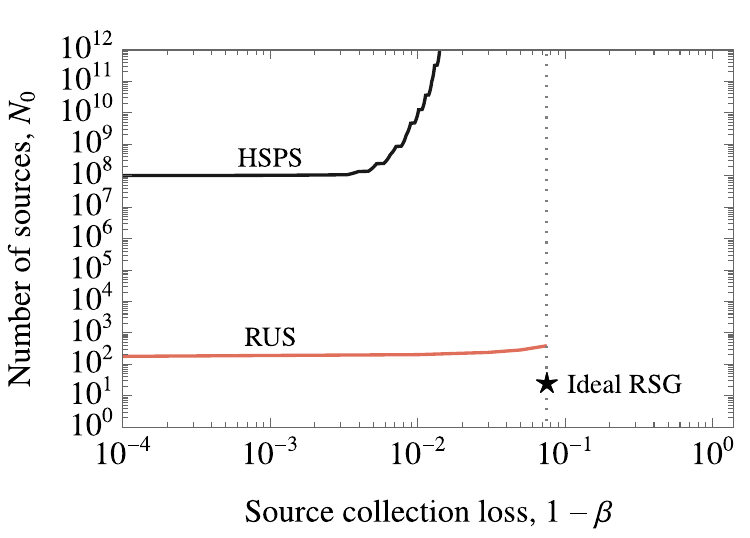}
    \caption{\textbf{Resource efficiency and maximal loss that each architecture can tolerate.} \textbf{a.} The regimes where fault-tolerant quantum computing is possible using the 24-photon Shor-encoded (2, 2) 6-ring resource state are indicated by shaded areas. The square points represent the hypothetical optimal implementations presented in Table \ref{tab:optimal}. The curves illustrate the simulated trade-off between resource efficiency and maximal loss per component when using MUX strategies to reach RSG success probability $p_\mathrm{rsg}=99.9\%$. For the HSPS, we assume a photon generation success probability of $p_s=5\%$, representing realistic improvements over current technology. For the RUS scheme, the solid line represents a single 12-source RSG while the dashed line represents the maximum resource efficiency achieved when using spatial multiplexing to reach $r=r_0$ at the cost of increased depth $D=2$. Both RUS curves consider only temporal resource sharing. \textbf{b.} The number of sources needed to reach $p_\mathrm{rsg}=99.9\%$ at $r=r_0=1$ GHz for the best (RUS) and worst (HSPS) architectures shown in panel a using resource sharing, spatial multiplexing, and a maximal loss per component of $x=0.1\%$. In both plots, we define the hypothetical ideal RSG architecture represented by the star point as an RSG that is able to produce resource states with perfect resource efficiency ($\eta_R=1$) and without photons passing through any optical components after being collected from source ($\eta_x=1$) so that the loss is purely determined by the collection efficiency $\beta$.}
    \label{fig:results}
\end{figure*}

The performance bounds computed in the previous section already provide an important order-of-magnitude comparison between the different approaches. However, as discussed in Section \ref{sec:tradeoff}, it is not straightforward to design an RSG architecture that approaches the maximum resource efficiency without also increasing optical depth, but there are some MUX strategies that help mitigate this trade-off.

Duplicate multiplexing can improve the trade-off by producing more than one copy of an intermediate state from the same register of sources to feed a single operation in the next stage. For instance, producing a 2-primate requires four identical photons. A single register of heralded sources can ensure that at least four photons are generated per clock cycle. This additional routing can be accomplished using the Spanke network discussed previously. However, in reality, the optical depth of the network will grow with an increasing amount of duplicate outputs.

Resource sharing \cite{bartolucci2021switch} is an extension of duplicate multiplexing where copies are grouped to feed multiple operations in the next stage. For example, a register of heralded sources can be connected to two modules, each producing a 2-primate state.  If four to eight photons are produced, one 2-primate state can be attempted, while eight or more allows for two attempts. This strategy reduces waste from multiplexing and improves resource efficiency toward the architecture's upper bound. Again, the multiplexing networks described earlier give appropriate flexibility for this task. But, as with duplicate multiplexing, the cost is having a more complicated switching network that will increase the optical depth with the number of photons being routed and thus decrease the maximal loss per component that is compatible with fault-tolerance.

To illustrate these trade-offs and provide a more practical comparison of the schemes, we simulate the resource efficiency while incorporating realistic multiplexing and resource-sharing strategies. 

To simulate photonic multiplexing used by the all-photonic and caterpillar schemes, we assume that the RSG is composed of a series of independent RSG modules that each use maximal resource sharing at all but the last stage. These independent RSGs are then multiplexed at the final stage to achieve the desired success probability of $p_\mathrm{rsg}=99.9\%$. For simplicity, we assume this final multiplexing stage can be accomplished in constant depth $3$ as before. The number of sources within each independent module and the resulting success probability is determined by the remaining available optical depth $D$ so that Eq. (\ref{eq:threshold}) is satisfied. For each MUX inside each module, we assume an optimistic sorting network of depth $\ceil{\log_2(n)}$ where $n$ is the number of states routed at the given stage. More details about the different figures and the optimization procedure are given in Appendix~\ref{app:Optitim}. Note that less optimal sorting might be achieved with a smaller depth, but this would not allow for maximal resource sharing.

For the RUS scheme, we performed a Monte Carlo simulation of the construction procedure that is illustrated in Figure \ref{fig:architectures}c (see Appendix~\ref{app:MonteCarlo} for details). For this scheme, we also consider two types of resource sharing. The first type is a temporal resource sharing technique that naturally occurs due to the spin memory. If a graph was not complete in time for the RSG cycle, the partially completed graph will be carried over into the next cycle, greatly increasing the probability of success. The second type is a spatial resource sharing technique that is possible due to having identical subgraphs in each stage of construction. In this case, the next stage of graph construction can commence as soon as any two subgraphs are complete, provided that the RUS gate connectivity can be reconfigured. To allow for this additional connectivity, and knowing that no delay lines are required to reconfigure the network due to the quantum memory, we assume a constant optical depth 3 on the RUS gate when using spatial resource sharing. Note that this does not increase the optical depth impacting the final photonic resource state, which remains $D=1$ due to the single switch.

Finally, we also consider using spatial multiplexing for the RUS scheme, where more than 12 sources are used to increase the success probability and RSG rate for the RUS scheme. The cost of this is to add an additional GMZI at the output of the RUS module to route photons into a predetermined spatial mode, although no delay lines are required.

In Figure \ref{fig:results}a, we show how four architectures compare in this practical setting using multiplexing and resource sharing techniques. Notably, in this practical setting, all four architectures fail to simultaneously reach the bounds on resource efficiency and maximal loss per component presented in Table \ref{tab:optimal}. However, the discrepancy is much greater for the all-photonic architectures compared to the caterpillar and RUS schemes. There is also a substantial drop in resource efficiency for the all-photonic architectures beyond a certain value of loss, which is caused by a rapid reduction in resource sharing in order to reduce the optical depth of each MUX stage to remain below the loss threshold of the resource state. On the other hand, the caterpillar scheme does not suffer a large trade-off because only one MUX stage uses resource sharing and it is already very shallow due to the relatively high success probability of performing two boosted fusion gates.

It is notable that the RUS scheme also suffers from a trade-off between the resource efficiency and the maximal loss per component, with a marked decrease in resource efficiency as the maximal loss per component approaches the upper bound. This is due to a reduced efficiency of the RUS gate, causing the module to restart some operations whenever a photon is lost (see Appendix~\ref{app:MonteCarlo} for more information on erasure errors). This increases the amount of internal clock cycles needed to ensure $p_\mathrm{rsg}$ reaches 99.9\%. However, when using spatial multiplexing, the resource efficiency can approach very close to the 20\% upper bound for the scheme even in the presence of significant loss. This latter result is a consequence of the fact that, for this scheme uniquely, resource sharing and multiplexing primarily impact photons used to entangle sources, rather than the final resource state itself.

Our results also show that, for a maximal loss per component of $x= 0.1\%$ and when considering spatial resource sharing, a multiplexed RUS module requires nearly six orders of magnitude fewer sources to achieve the same rate of generation as an HSPS with $p_s=5\%$ (see Figure \ref{fig:results}b). Improving $p_s$ to at most $25\%$ can reduce the resource overhead of the all-photonic architecture by factor of around five, but this surprisingly provides very little improvement to the loss tolerance. Moreover, due to the shallow optical depth, the RUS scheme can achieve this resource advantage using sources with an order of magnitude higher collection loss.

\section{Discussion}
\label{sec:discussion}

\paragraph{Natural constraints on RSG rates.}
Interpreting $\eta_R$ in the context of a physical platform is important to ensure physical suitability. When using an RSG architecture to perform fault-tolerant computations, the rate $r_0$ is constrained by classical electronics due to feedforward operations and detector reset times (estimated at most $\sim$1~GHz for dedicated electronics) while the RSG rate $r$ is limited by the error correction cycle, which is dominated by the decoding time (estimated at $\sim$1~MHz \cite{battistel2023real}). If an RSG outpaces decoding, interleaving can be used to reduce the number of RSGs needed for the overall FTQC architecture \cite{bombin2021interleaving, kim2022fault}. Thus, like temporal multiplexing, this is a strategy that can be used to trade rate to reduce the physical source overhead. However, interleaving cannot be used to increase the effective RSG efficiency above the upper bound, and it also reduces the acceptable maximal loss per component due to memories and routing.

\paragraph{Implication for industrial-scale applications.} Using low-loss fibers and interleaving \cite{bombin2021interleaving}, it has been estimated in Ref.~\cite{kim2022fault} that one RSG can effectively feed multiple logical qubits in the fusion network, enabling industrial-scale applications with between $10^3$ and $10^5$ RSGs. This benefit applies broadly to FBQC architectures, regardless of the specific RSG implementation. Thus, with a minimum of 12 sources per RSG, an FBQC architecture using RUS-based RSGs could achieve industrial-scale fault-tolerant computation with as few as 12000 sources. This is in stark contrast to a surface code architecture implemented with superconducting circuits, which would require over a million physical qubits, and to the all-photonic FBQC architecture using HSPSs studied in this work, which we estimate would require upwards of one billion sources---around five orders of magnitude more than a RUS-based architecture. A hybrid spin-photonic approach can therefore reduce the resource overhead of fault-tolerant quantum computation by multiple orders of magnitude compared to both all-matter and all-photonic platforms. Such a major reduction is expected to translate directly into a substantially lower fabrication and operation costs.

\paragraph{Impact of future architecture and component improvements.}
This study compares high-performing schemes using best-known approaches, though none of the explored architectures are expected to be optimal and may be improved in the future, perhaps by considering specific hardware details or different resource states.

For instance, it is possible to design photonic schemes using fewer stages of multiplexing, provided that switching networks are highly transmissive but where the majority of loss occurs due to the photonic memory needed to reconfigure the switching network. It may also be possible to determine a more optimal division of the construction into intermediate states \cite{lee2023graph}, better use of auxiliary resources \cite{pankovich2024flexible}, or the recycling of photonic states after failed operations \cite{browne2005resource} at the cost of increased loss and error.

On a similar note, the 12-source RUS scheme was designed to minimize the optical depth seen by the resource state. However, there are possibly more resource-efficient schemes that combine the caterpillar graph state generation with RUS gates to reach even better upper bounds on resource efficiency. The consequence will be to use at least one stage of photonic multiplexing, which will increase the minimum optical depth. But, in practice, when accounting for the trade-off between resource efficiency and loss tolerance, such a scheme could outperform the 12-source scheme in both metrics in some parameter regimes.

Our analysis provides a foundation for extending comparisons to other graph codes and construction schemes, enabling a broader mapping of RSG architectures in terms of resource efficiency and optical depth for different source types. While we have focused on critical metrics—component count, entanglement success, and loss tolerance—that have a profound impact, refined comparisons must also incorporate other errors that arise during state construction and extend the analysis to full error correction simulations in order to determine fault-tolerance thresholds. Although accurate threshold simulations are beyond the scope of this work, in Appendix~\ref{app:twoqubitinteractions}, we estimate the number of effective photon-photon and spin-photon interactions required by each scheme. This analysis reveals that gains in resource efficiency are generally accompanied by an increased reliance on spin-photon interactions. Notably, while the balance between photon-photon and spin-photon operations varies across schemes, the total number of effective two-qubit interactions remains broadly comparable. This suggests that, at a first approximation, all four schemes may exhibit similar overall sensitivity to noise. However, capturing the true performance impact will require detailed modeling of physically motivated error channels and full-scale simulations of the noisy construction circuits.

\paragraph{Suitability and limits of current state-of-the-art sources for the proposed RSG architectures.}
In addition to the optical depth, it is crucial to consider the efficiency of the source itself. For heralded sources, free-space implementations have shown efficiency values up to 94\% \cite{zhong201812} before lossy collection into a fiber. On-chip implementations necessary to mass fabricate sources can produce high purity photons, but reaching high source efficiency remains a challenge with values around 26\% \cite{alexander2024manufacturable}. Deterministic sources have improved significantly over the past decade, with micropillar cavities consistently demonstrating efficiency values above $50\%$ before lossy collection into a fiber \cite{thomas2021bright, maring2024versatile}. On-chip sources have reached efficiencies up to 84\% \cite{uppu2020scalable}, but efficient coupling to low-loss waveguides or fibers has yet to be demonstrated. State-of-the-art values are obtained using a tunable open cavity design. This latter approach has demonstrated the highest fibered efficiencies, with $57\%$ in 2020 \cite{tomm2021bright} and $71\%$ in 2023 \cite{ding2023high}. However, the manufacturability and stability of open cavity designs remains challenging.

In addition to photon collection efficiency, photons that are highly pure and indistinguishable are necessary to ensure high-fidelity fusion and RUS gates. Currently, HSPS hold the record for indistinguishability of up to 99.5\% for photons from distinct sources, but come with a multi-photon error of 0.7\% (integrated autocorrelation of $g^{(2)}(0)=0.35\%$) \cite{alexander2024manufacturable}. This also comes along with a low success probability (est. $p_s<1\%$) limited by spectral filters that cause $20\%$ loss \cite{alexander2024manufacturable} and hence prevent efficient photon-number resolved heralding that suppresses multi-photon emission produced at higher success probabilities. The state-of-the-art indistinguishability between photons emitted from distinct quantum dots in bulk material is 93.0\% \cite{zhai2022quantum}, limited by a combination of phonon-induced dephasing and charge noise. Further improvements are expected by combining the high-quality fabrication of Ref.~\cite{zhai2022quantum} with cavity-enhanced emission \cite{grange2017reducing}, which suppresses the phonon sideband and shortens the emission lifetime—reducing both dephasing of the zero-phonon line and charge noise to enable indistinguishability above 99\%. Quantum dots also consistently provide sub-1\% multi-photon errors without compromising other metrics \cite{maring2024versatile}, and they hold the record for multi-photon errors as low as 0.02\% \cite{hanschke2018quantum}. 

On-chip optical components are also reaching the high levels of quality needed to implement linear-optical processing. Twenty-mode universal SiN chips are commercially available that reach $>97\%$ transformation fidelity \cite{taballione202320}. Using machine learning characterization, universal interferometers can also reach fidelity exceeding 99.77\% \cite{fyrillas2023scalable}. Smaller specialized integrated circuits show even higher fidelity, with a standard Mach-Zehnder interferometer reaching a fidelity up to 99.996\% when characterized with coherent light \cite{wilkes201660, alexander2024manufacturable}. Component losses are also improving, with SiN integrated beam splitters and mode crossings reaching 0.012\% and 0.037\%, respectively \cite{alexander2024manufacturable}. The components that must still be improved are the integrated PNR detectors, the chip-to-fiber interface and high-bandwidth phase shifters. The last two currently incur a loss of 1.3\% and 2.3\%, respectively~\cite{alexander2024manufacturable}, which exceed the maximum loss per component estimated for all but the RUS-based architecture. 

Solid-state QD sources with a matter qubit degree of freedom have recently demonstrated the generation of small entangled states \cite{appel2022entangling,coste2023high, meng2024deterministic, su2024continuous, huet2024deterministic}. These early demonstrations use devices with spin coherence times on the order of nanoseconds. Given that photon emission from cavity-enhanced QDs occurs on a 100~ps timescale, the repetition period can be reduced to 1~ns. This allows for large entangled states to be produced even with seemingly short coherence times. For example, it is predicted that 14-photon graph states with $75\%$ fidelity (less than 2\% error per photon) could be achieved with only minor improvements to current devices (see supplementary material of  Ref.~\cite{huet2024deterministic}). Since state-of-the-art spin coherence times of up to 100 $\mu$s have already been achieved with QDs using dynamical decoupling \cite{zaporski2023ideal}, we also anticipate that incorporating a mild amount of dynamical decoupling into the generation protocol of Ref.~\cite{huet2024deterministic} will eliminate spin decoherence as a dominant source of error.

Dynamical decoupling is also particularly well-suited for the RUS scheme, where the sources must idle between RUS attempts during feedforward, after completing all their edges, or while waiting for the emission of the resource state. Thus, standard dynamical decoupling sequences could be directly adopted, and $100~\mu$s is more than sufficient to construct a 24-photon resource state with up to $99.95\%$ fidelity using the RUS scheme, even when limited by current feedforward times of 23 ns \cite{thiele2024cryogenic}. 

With extended coherence times and improved spin-photon entanglement quality, resource state fidelity will be primarily limited by single-qubit gates, which can already reach 99.3\% \cite{zaporski2023ideal} and are expected to improve substantially in an industrial setting, in addition to the RUS gate fidelity, which depends primarily on the purity and indistinguishability of the emitted photons.

\section{Conclusion}
\label{sec:outlook}

The main strengths of photonic quantum computers are modularity, scalability, and flexibility. It is often said that the main drawback is the resource overhead. However, we demonstrated that this overhead practically vanishes when using deterministic single-photon sources with a single matter qubit degree of freedom.

The RUS module architecture builds on these strengths by harnessing a key advantage of material platforms---deterministic entangling gates. This resource-efficient approach allows high-performance sources to be selected and connected to a RUS module, regardless of their location on a chip. Defective sources and components can be easily replaced or upgraded. Additionally, the module can be reconfigured after fabrication, either by rerouting fibers or programming a photonic integrated circuit, to generate different resource states depending on the task.

Our results, combined with recent tailored architectures beyond the fusion-based paradigm leveraging RUS gates and measurements \cite{de2024spin, paesani2023high, hilaire2024enhanced}, highlight the untapped potential of hybrid platforms. These advancements open critical avenues for developing more efficient fault-tolerant photonic architectures, accelerating progress toward utility-scale quantum computation while minimizing resource costs.
\\
\section*{Acknowledgments}

The authors thank Gr\'egoire de Gliniasty, Katia Hakem, Emilio Annoni, and Niccolo Somaschi for fruitful discussions; Rawad Mezher, Nicolas Maring, Jean Senellart, Thomas Volz, and Andrew White for valuable feedback on the manuscript. This work has been co-funded by the
Plan France 2030 through the PROQCIMA program, the
TUF-TOPIQC program, the OECQ i-démo project, and the ANR projects ANR-22-PETQ-0006 (NISQ2LSQ) and
ANR-22-PETQ-0013 (Oqulus). This work has also been
co-funded by the Horizon-CL4 program under the grant
agreement 101135288 for the EPIQUE project. 

\appendix 

\section{List of symbols and figures of merit}
\label{sec:recap-table}
For clarity and ease of reading, the following table summarizes the main symbols and figures of merit used throughout this work, together with their definition and relation to other parameters.

\begin{table}[h!]
\centering
\renewcommand{\arraystretch}{1.3}
\begin{tabular}{@{} l >{\raggedright\arraybackslash}p{0.20\textwidth} >{\centering\arraybackslash}p{0.20\textwidth} @{}}
\hline
\textbf{Symbol} & \textbf{Description} & \makecell{\textbf{Relation to}\\\textbf{other quantities}} \\
\hline
$p_s$   & SPS success probability     &  \\
$\eta_H$  & SPS heralding efficiency      & \\
$p_{\textrm{pair}}$   & SPS photon pair  creation probability      & $p_s= \eta_H p_{\textrm{pair}}$ \\
$N$             & Number of photons in the final resource state        &  \\
$p_\mathrm{rsg}$             & RSG succes probability  &  \\
$r$             & RSG clock rate        &  \\
$r_0$             & RSG internal clock rate (laser pump rate)     &  \\
$\eta_R$             & Resource efficiency   & $\eta_R = Nr/(N_0r_0)$ \\
$N_\text{avg}$  & Average number of photons to produce one resource state   & \\
$\eta_\text{mux}$   & Multiplexing efficiency     & $\eta_\text{mux} = N_\text{avg}r/(p_sN_0r_0)$ \\
$\tau_\text{avg}$   & Average number of internal clock cycles for the RUS module  &  \\
$\tilde{\eta}_R$  & Resource efficiency bound  & $\tilde{\eta}_R= N/(N_0\tau_\text{avg})$ (RUS)  $\tilde{\eta}_R= p_sN/N_\text{avg}$  (others)\\
$\beta$             & Source collection efficiency      &  \\
$\eta_x$             & Network transmission efficiency      &  \\
$\eta_T$             & RSG total transmission efficiency      & $\eta_T=\beta \eta_x$ \\
$x$             & Loss per component      & \\
$D$             & Optical depth of the network   & $\eta_x=(1-x)^D$ \\
$N_\text{mux}$ & Number of multiplexing stages & \\
$D_\text{GMZI}$ & GMZI depth & \\
$D_\text{coupl.}$ & Coupling depth & \\
$D_\text{s}$ & Optical depth per spatial MUX stage & \\
$D_\text{b}$ & Optical depth per spatio-temporal MUX stage (de Bruijn)& \\
\end{tabular}
\caption{Summary of symbols and figures of merit used in this work.}
\label{tab:symbols}
\end{table}

\section{Reminders on graph states and fusion gates}
\label{sec:graph-prelim}

\subsection{Graph States}
\label{ssec:graphstates}
A graph state is a quantum state defined by a graph $G = (V, E)$ such that:
\begin{equation}
    \ket{G} = \prod_{(v,w) \in E} \mathsf{CZ}_{v,w} \bigotimes_{v \in V} \ket{+}_v.
\end{equation}
Let $\ket{\downarrow}$ and $\ket{\uparrow}$ be the computational basis states of the matter qubit degree of freedom of an SPS. We assume that, after excitation, this SPS emits a photon whose polarisation is correlated to the spin state via the following emission operator:
\begin{equation}
    E_{\rm qe} = \ket{\downarrow, L}\bra{\downarrow} + \ket{\uparrow, R} \bra{\uparrow},
    \label{eq:quantum_emitter}
\end{equation}
where $\ket{L}, \ket{R}$ denote respectively left and right polarised photon states.

\begin{figure}
\centering
\includegraphics[width=2\columnwidth/3]{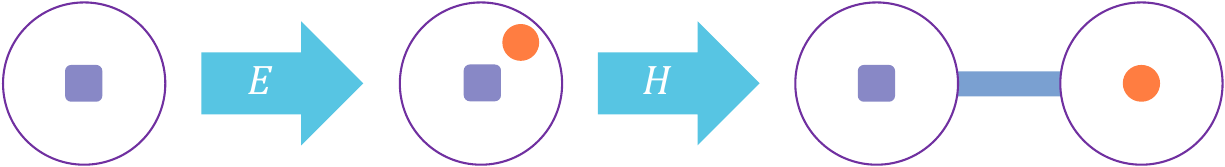}
\caption{Basic linear cluster generation process. Starting with the emitter's qubit in the $\ket{+}_{qe}$ state (small purple square), a Bell state $\ket{\Psi}$ is created via the emission of a photon (small orange circle) and then transformed into a cluster state with a Hadamard gate applied to the SPS's qubit. The middle circle containing two elements represents the Bell state as a redundantly-encoded $\ket{+}$ state.}
\label{fig:basic-em}
\end{figure}

This is equivalent to applying a $\mathsf{CNOT}$ between the quantum emitter and a photon in the state $\ket{0} = \ket{L}$. Applying this operator to a matter qubit in the $\ket{+}$ state yields a Bell state, and repeating this operation produces a GHZ state, which can be seen as a redundantly-encoded $\ket{+}$ state. Applying a Hadamard gate to the matter qubit after the first emission instead ``pushes'' the matter qubit into another node (see~\cite{lindner2009proposal,hilaire2023near}), yielding the state $\mathsf{CZ}_{1,2}\ket{+}\ket{+}$. This process is depicted in Figure \ref{fig:basic-em}.

\begin{figure}
\centering
\includegraphics[width=2\columnwidth/3]{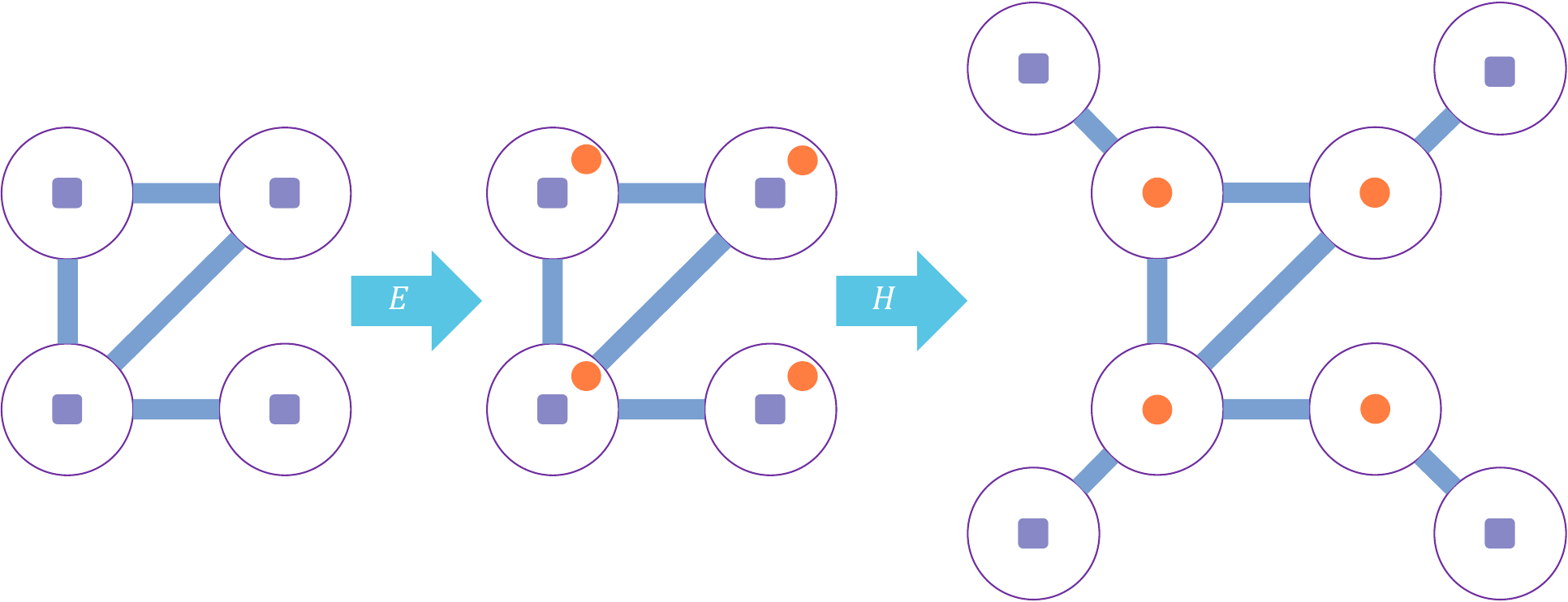}
\caption{Starting from a graph state on the emitter qubits, emitting one photon generates a redundantly-encoded graph state, and applying $\mathsf{H}$ on the emitters ``pushes them out'' of their vertex into another which is only linked to their emitted photon.}
\label{fig:graph-transfer}
\end{figure}

More generally, given a graph state $\ket{G}$ between $n$ such SPSs, having each SPS emit one photon via this process will first yield the same graph state but with each vertex being redundantly encoded. Then, applying a Hadamard gate to the qubit of each SPS will transfer the graph to the photons, with each SPS only linked to its emitted photon. An example is shown in Figure~\ref{fig:graph-transfer}.

\subsection{Fusion gates}

We now explain how fusion gates~\cite{browne2005resource} are performed and their effect on graph states.

A type-I fusion is a parity measurement, performed by the circuit in Figure~\ref{fig:fusion1}. If a single photon is detected, which happens with probability $50\%$, then the state is transformed by the Kraus operators $\mathsf{Z}^b(\ket{H}\bra{HH} + \ket{V}\bra{VV})/\sqrt{2}$, where the bit $b$ is equal to $1$ if the photon was measured in the vertical polarization. The effect on a graph state is that the qubit encoded in the unmeasured photon inherits all the links that the measured qubit used to have, i.e.~the two nodes have been fused together. A fusion failure, indicated by measuring either 0 or 2 photons, is equivalent to performing a $\mathsf{Z}$ measurement on both nodes, removing them and all their edges from the graph.

The type-II fusion that we use, on the other hand, is a rotated Bell measurement $\langle\mathsf{XZ}, \mathsf{ZX}\rangle$ on the two qubits. It is performed by the circuit in Figure~\ref{fig:fusion2}. It is successful if one photon is detected in the upper two modes and another is detected in the lower two modes, which happens again with probability $50\%$. Applying a type-II fusion to two nodes $v_1, v_2$ of the graph that were previously not linked has the same effect as applying a type-I fusion to $v_1$ and $v_2$, thus fusing the two nodes into a single node $v$, and then performing a $\mathsf{Y}$ measurement on $v$, which complements its neighborhood and then removes it from the graph. Failing the fusion corresponds to measuring $\mathsf{Z}$ on both qubits. Furthermore, using an additional Bell pair~\cite{grice2011arbitrarily} or four additional single photons~\cite{ewert20143}, it is possible to boost type-II fusions so that their success probability is $75\%$ instead of $50\%$.

\subsection{RUS CZ gate}

\begin{figure*}
\centering
\begin{subfigure}[t]{0.3\textwidth}
\includegraphics[width=\linewidth]{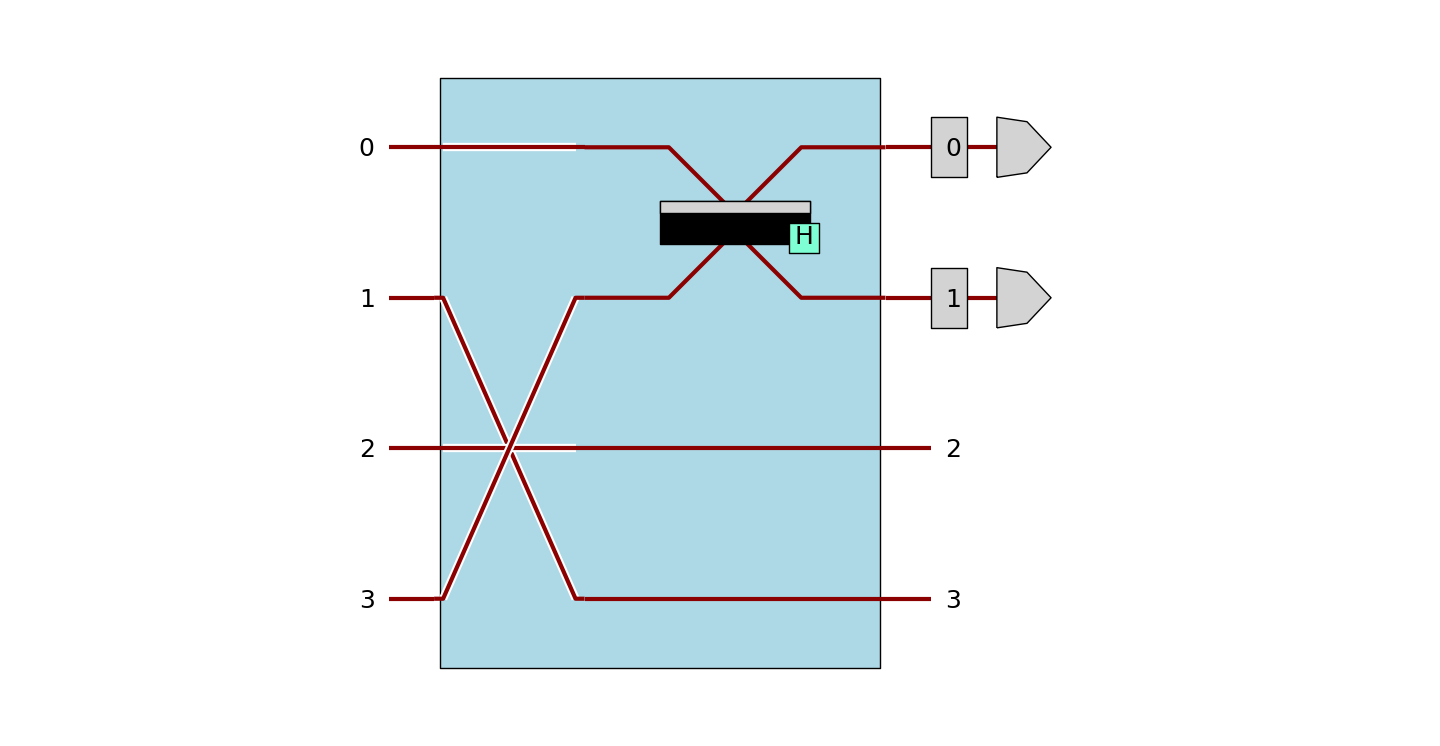}
\caption{Type I Fusion.}
\label{fig:fusion1}
\end{subfigure}
\begin{subfigure}[t]{0.3\textwidth}
\includegraphics[width=\linewidth]{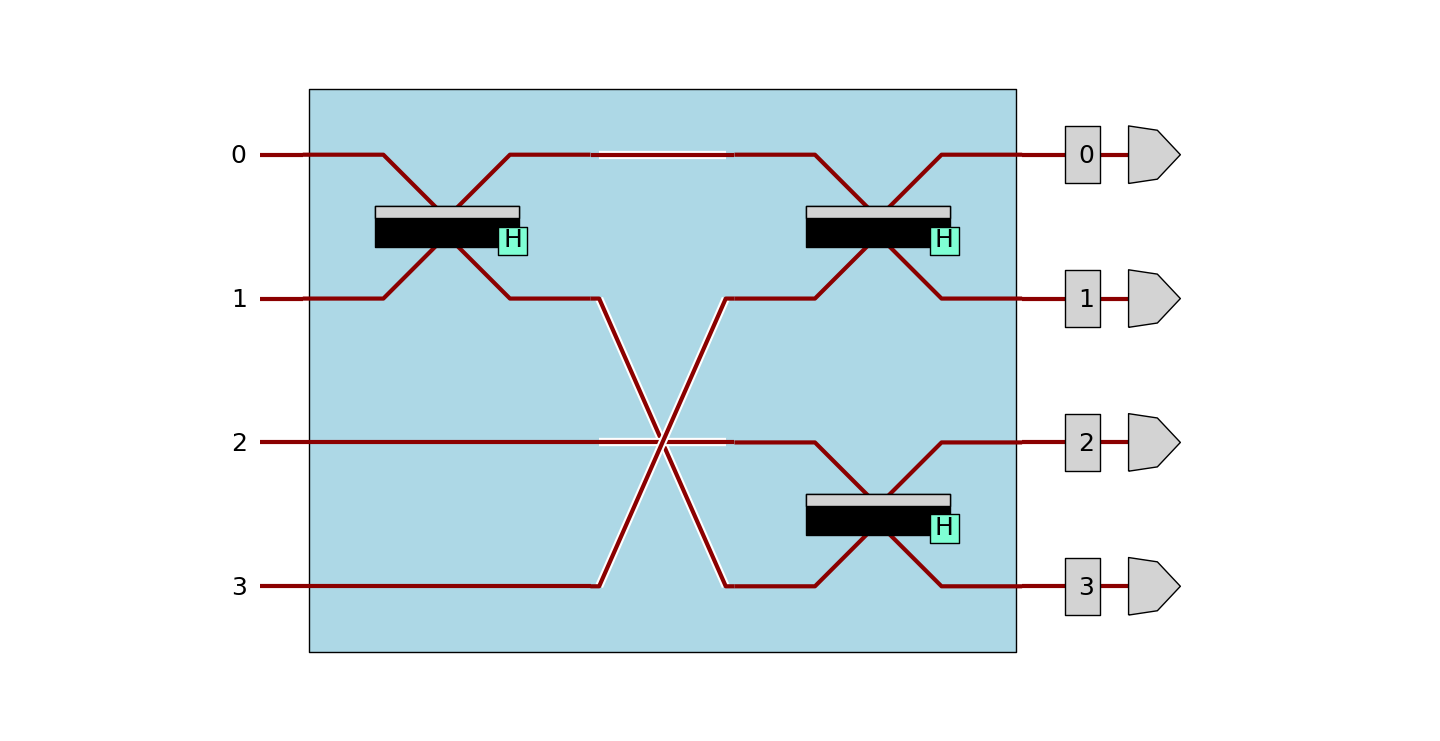}
\caption{Type 2 Fusion.}
\label{fig:fusion2}
\end{subfigure}
\begin{subfigure}[t]{0.3\textwidth}
\includegraphics[width=\linewidth]{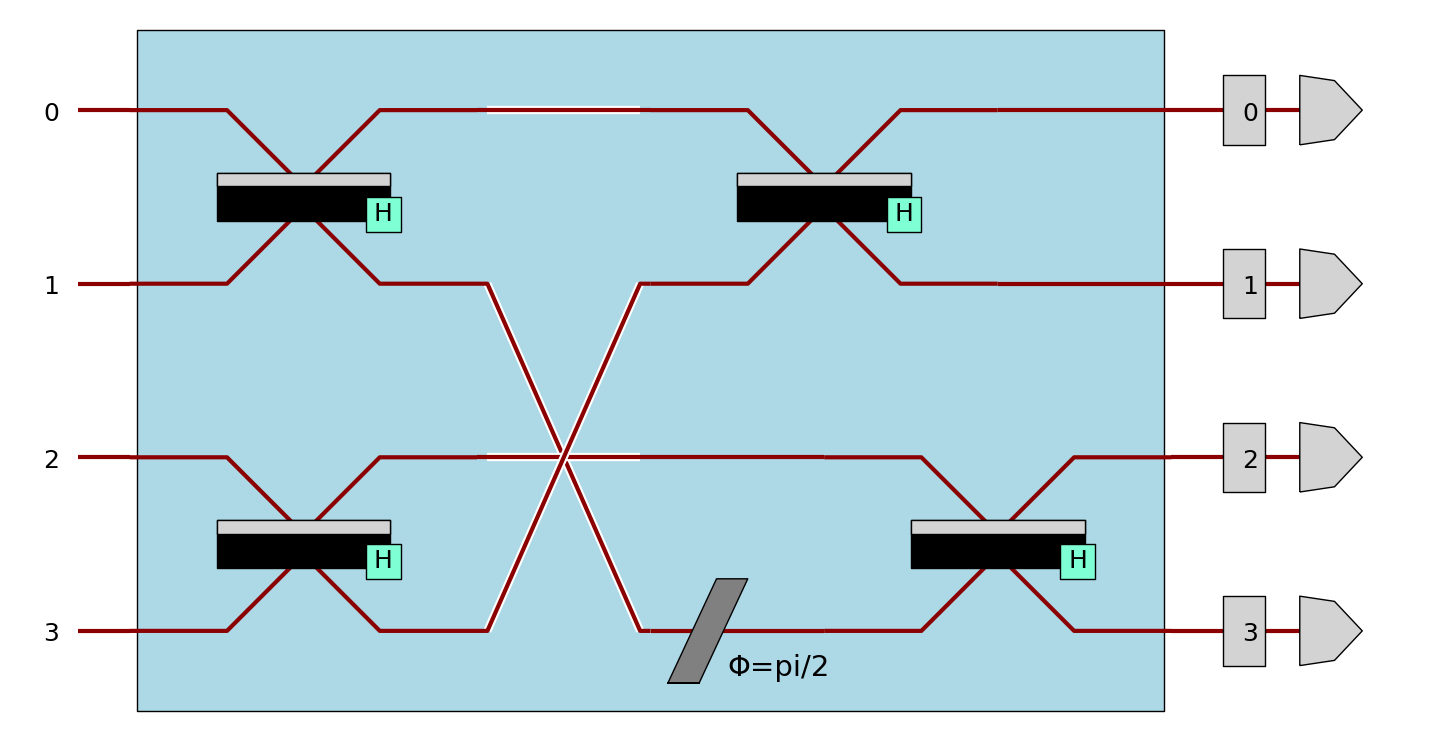}
\caption{RUS gates.}
\label{fig:rus}
\end{subfigure}
\caption{Linear-optical circuits for fusion gates and for the RUS gate used in the proposed architectures, assuming the qubits are encoded in dual spatial rail. The first two input modes correspond to the qubit of one graph while the last two correspond to the other graph's qubit. The horizontal bars are 50:50 beamsplitters using the Hadamard convention, the slanted gray bar in panel (c) is a phase-shifter with phase $\pi/2$, the detectors are able to resolve up to 2 photons at least.}
\label{fig:fusion}
\end{figure*}

One attempt of our RUS $\mathsf{CZ}$ gate, as described in Ref.~\cite{de2024spin}, consists of the emission operator Eq.~\ref{eq:quantum_emitter} followed by a linear-optical joint measurement illustrated in Figure~\ref{fig:rus}. More precisely, applying the emission operator is equivalent to instantiating the photonic qubit in the $\ket{0}$ state and applying a $\mathsf{CNOT}$ gate between the spin qubit (as the control) and the photonic qubit (as the target). The RUS gate circuit in Figure~\ref{fig:rus} implements a $\langle\mathsf{XX}, \mathsf{ZY}\rangle$ stabilizer measurement on the photonic qubits in case of success. Since these are each entangled with their respective spin qubits via the emission operator, this corresponds to the teleportation of an entangling $\mathsf{CZ}$ gate to the spin qubits (up to local corrections). A failure corresponds to measuring both photons in $\mathsf{X}$ so that the identity gate is teleported to the spin qubits up to a phase correction, enabling repeated attempts.

The circuit in  Figure~\ref{fig:rus} that implements the entangling measurement on the photonic qubits is equivalent to a rotated type-II fusion gate. However, unlike fusion operations, RUS gates are not restricted to graph states only. The RUS CZ protocol will teleport a CZ gate to spins initialized in any state. Since both the RUS $\mathsf{CZ}$ gate and type-II fusion gates measure both photonic qubits, loss errors can be detected and used to signal adaptive reconstruction of lost graph edges or a hard reset on the graph construction.

\section{Multi-photon emission for probabilistic sources}
\label{appendix_g2}

\begin{figure}
    \centering
    \includegraphics[width=\linewidth]{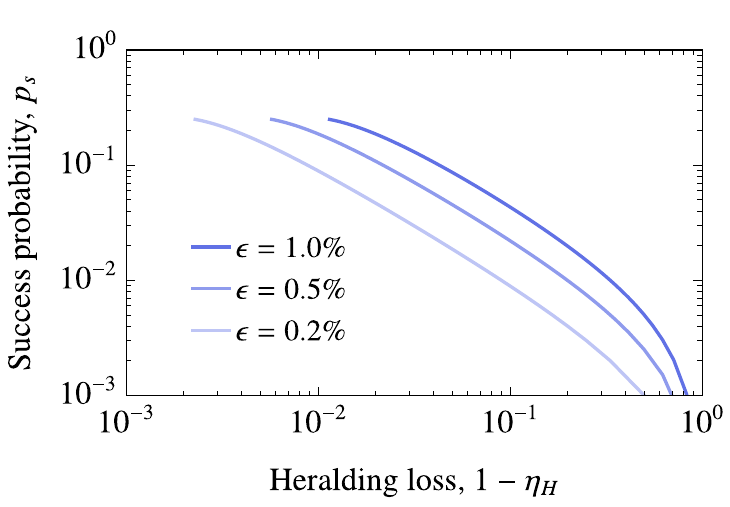}
    \caption{The success probability $p_s$ that can be achieved for a given heralding probability $\eta_H$ when constraining the multi-photon error probability $\epsilon$.}
    \label{fig:g2sfwm}
\end{figure}

A basic model for the photonic state produced by a probabilistic photon source \cite{braunstein2005quantum} is
\begin{equation}
\label{eq:sfwm}
    \ket{\psi} = \sqrt{1-\lambda^2}\sum_{n=0}^\infty \lambda^n\ket{n,n}
\end{equation}
where $|\lambda|^2 = \tanh^2(a)$ and $a$ is a dimensionless parameter that is proportional to the pumping laser amplitude. In the lossless scenario, a threshold detector can eliminate the possibility of $n=0$, thus heralding the presence of at least one single photon with the probability $p_s=|\lambda|^2$. However, the probability of producing more than one photon is $|\lambda|^4$, and hence the probability of producing more than one photon given that at least one photon was detected is $\epsilon=|\lambda|^2=p_s$. Therefore the pair generation probability and the multi-photon noise are tied together.

One solution to this problem is to resolve the number of photons in the idler signal. For perfect detection efficiency and fidelity, this eliminates multi-photon errors entirely because they always correspond to more than one photon in the idler, which can be rejected. In this case $p_s=|\lambda|^2(1-|\lambda|^2)$ and $\epsilon=0$.

In reality, the idler signal is often passed through filters that shape the spectrum of the photon so that only high-quality single photons are heralded in the signal. This causes loss on the idler prior to detection, which reduces the heralding probability $\eta_H$, and in turn allows for multi-photon errors to leak into the signal. In addition, the detector itself must have a very high detection efficiency and fidelity, or else this critical measurement may not provide enough accurate information to ensure that only proper single photons are heralded in the signal.

The quality of photon-number heralded single photons can be modelled by first applying a loss channel to the idler mode in Eq. $\ref{eq:sfwm}$. The loss channel is $\Lambda = e^{-\log(\eta_H)\mathcal{D}}$, where $\mathcal{D}(\rho)=\hat{a}\rho\hat{a}^\dagger-\hat{a}^\dagger\hat{a}\rho/2-\rho\hat{a}^\dagger\hat{a}/2$, $\hat{a}$ is the annihilation operator of the idler mode, and $\rho=\ket{\psi}\bra{\psi}$ is the full state of the idler (I) and the signal (S). Then, the state is projected onto the single-photon subspace of the idler $\rho_\text{S} = \text{Tr}_\text{I}\left(\ket{1}\bra{1}_\text{I}\Lambda(\rho)\right)$. The success probability is thus $p_s=\text{Tr}_\text{S}\left(\rho_\text{S}\right)$ and the multi-photon error probability is $\epsilon = 1 - (\bra{0}\rho_\text{S}\ket{0} + \bra{1}\rho_\text{S}\ket{1})/p$.

Figure \ref{fig:g2sfwm} shows the success probability $p_s$ that can be reached for a given maximum amount of loss on the idler when constrained to achieve a multi-photon error of $\epsilon=1.0\%, 0.5\%$, and $0.2\%$. This simulation assumes all other processes are perfect, including the photon-number resolving fidelity. For a $0.5\%$ error, which would give an integrated intensity autocorrelation of $g^{(2)}\sim 1\%$, the minimum detection efficiency needed to reach $p_s=25\%$ is $\eta_H=99.4\%$. Requiring an even lower error rate of $\epsilon=0.2\%$ implies a detection efficiency of $\eta_H=99.8\%$ is needed and so the filters used to reduce spectral noise must not cause more than $0.2\%$ loss.

\section{Derivation of ${N_\text{avg}, i}$}
\label{sec:appNAvg}

At each stage $i$ we aim at producing a specific amount of resource states which will be used at a later stage. We want to compute the average number of photons $N_{\text{avg},i}$ needed to produce such states. We proceed recursively. To produce one state at stage $i$: 
\begin{itemize}
    \item one first needs to produce $c_{i-1}$ states at stage $i-1$;
    \item one then needs to input these $c_{i-1}$ states into a network of probabilistic linear optical operations to produce one desired state. This operation succeeds with probability $p_i$.
\end{itemize}
This procedure is repeated $c_i$ times to obtain the $c_i$ desired resource states.

Let $X$ be the random variable associated to the number of photons needed to produce $c_{i-1}$ states at stage $i-1$, let $Y$ be the random variable associated to the number of attempts to produce one state at stage $i$ given that $c_{i-1}$ states were produced at stage $i-1$, and let $Z$ be the random variable associated to the number of photons needed to produce $c_i$ states at stage $i$.
Then $E[X] = N_{avg,i-1}$, $E[Z] = N_{avg,i}$ is the quantity we want to compute, and $Y$ follows a geometric distribution, because each attempt is independent and succeeds with probability $p_i$, so
\begin{equation}
     P(Y=k) = (1-p_i)^{k-1} p_i.
\end{equation}

During one attempt to produce one state, we need $y \in Y$ trials, and a total of $x+a_i$ photons per trial, where $x \in X$ and $a_i$ the number of auxiliary photons used to boost the fusions. Repeating this $c_i$ times, the total number of photons $z \in Z$ needed at stage $i$ is: 
\begin{equation}
     z = \sum_{j=1}^{c_i} \sum_{k=1}^{y} \left( x_{kj} +a_i \right)
\end{equation}
where $x_{kj} \in X$ are independent realizations of $X$. In other words: 
\begin{equation}
    Z = \sum_{j=1}^{c_i} \sum_{k=1}^{Y} \left( X + a_i \right). 
\end{equation} 
Since $X$ and $Y$ are independent, the expectation value of $Z$ is equal to: 
\begin{equation}
 E[Z] = c_i \cdot E[Y] \cdot (E[X] + a_i),
\end{equation} 
which corresponds to the expression in the main text.

\section{Boosted fusions}
\label{apx:bounds}

\subsection{All-photonic}
\label{apx:allphotonic}

The proposed all-photonic approach to building the 24-photon resource state is broken down into six stages. Each stage requires a number of copies of the successfully-produced state from the previous stage and has its own success probability. The number of copies and success probability associated with each stage is summarized in Table \ref{tab:allphotonicsummary} of the main text. Here, we consider that the last three stages use a boosted fusion probability of $1-1/2^b$ rather than 3/4.
 
For $b=1$, we compute an average number of photons of $N_\text{avg}=237327$. The $b=2$ case is computed in the main text using four additional single photons per fusion, which gives $N_\text{avg}=21170$. For $b=3$, we assume that an additional 4-GHZ state is used for each fusion gate, which each require $309$ photons on average to produce. Taking into account this cost, the total average number of photons only decreases to $N_\text{avg}=19585$. We judge that saving ~7\% on the number of photons on average does not justify the decrease in gate success probability due to the significant chance that at least one of the auxiliary photons is lost, nor the increased complexity of the integrated circuit and routing of additional GHZ states between stages.

\subsection{Caterpillar source}
\label{apx:caterpillar}
The caterpillar source deterministically produces a single seed state composed of 14 photons. Four of those photons are used to perform two type II fusion gates. Three other photons are measured, but this is a deterministic operation in the lossless scenario. Three copies of these states are then fused into the final resource state using three more type II fusion gates. 

Using the values from Table \ref{tab:caterpillarsummary} in the main text, and instead assuming $b=1$ for all fusion gates, we compute the average number of photons consumed to be $N_\text{avg}=1344$. For $b=2$, as computed in the main text by assuming an additional Bell pair is used, we have $N_\text{avg}=242$. For $b=3$, assuming that an additional 4-GHZ state is produced by the same source to boost each fusion gate, the average number of photons decreases slightly to $N_\text{avg}=147$. This improvement is relatively good compared to the all-photonic architecture, representing a 40\% decrease in the number of photons consumed. However, we again judge that the gain is relatively small compared to moving from $b=1$ to $b=2$ and thus will not justify the increased optical losses and complexity.

\section{Simulation of a perfect resource sharing model} \label{sec:simulation_perfect_sharing}

Given a general scheme for resource state generation, this section describes one way to estimate the success probability $p_\mathrm{rsg}$ of having at least one resource state at the output. We assume perfect sharing i.e, all states at any stage will be used in the computation because we can freely route them around with a sorting network.

\begin{figure*}
    \centering
    \includegraphics[width=\linewidth]{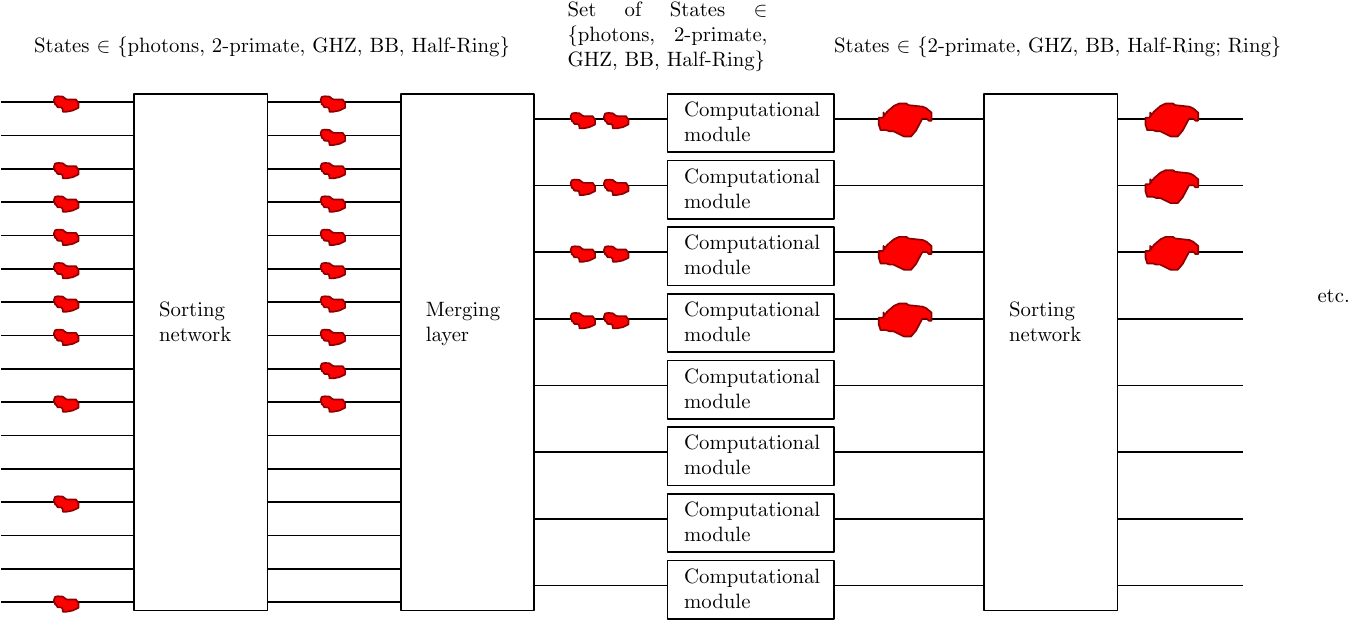}
    \caption{How the computation of the all-photonic scheme with photon sources or caterpillar sources is handled in our simulation. Three stages are repeated to produce more and more complex resource states. After each stage, we update the probability distribution $P(X=j)$ of having $j$ resource states at the end of the stage. We illustrate the possible intermediate resource states for the all-photonic example with photon sources. BB stands for ``Building block", see Fig.~\ref{fig:architectures}a.}
    \label{fig:resource_sharing}
\end{figure*}

As depicted in Figure~\ref{fig:resource_sharing}, the scheme is divided into a series of three stages: 
\begin{itemize}
    \item a sorting stage where intermediate states are gathered together in neighbor modes, 
    \item a merging stage where intermediate states are ``merged" into sets of $c$ copies, where $c$ is the number of copies needed to proceed at the next stage, here we refer as the basic resource a ``set of states",
    \item finally, a computation stage that takes as input $c$ copies of a state and tries to produce a bigger state with probability $p$.
\end{itemize}

Throughout the simulation, we keep track of the probability distribution $P(X=j)$ of having $j$ states/sets of states during the current stage. Then the probability distribution is modified depending on the stage. Once we have the probability distribution $P_{\text{final}}$ of the final stage, the success probability of the whole process is given by $p_\mathrm{rsg} = 1 - P_{\text{final}}(X=0)$.

\subsubsection{Sorting stage}

The sorting stage does not modify the probability distribution. It moves the states around, but overall the numbers of states is unchanged. 

\subsubsection{Merging stage}

In the merging stage, we gather states into sets of $c$ states. So we have one set of states if the number of states is between $c$ and $2c-1$, we have two sets if the  number of states is between $2c$ and $3c-1$, etc. Overall, if $P'$ is the new probability distribution and $P$ the current one, the following holds:

\begin{equation}
    P'(X=j) = \sum_{i=0}^{c-1} P(X=jc + i).
\end{equation} 

\subsubsection{Computational stage}

At this stage we have sets of states that have to be processed into a bigger state. This operation succeeds with probability $p$. Overall, the probability of having $j$ new states after the stage is the sum of the scenarios ``I have $k \geq j$ sets of states and $j$ of them have succeeded". In other words, if $m$ is the maximum number of states we can expect at this stage, then:

\begin{equation}
    P'(X=j) = \sum_{k=j}^{m} P(X=k) \cdot \binom{k}{j} \cdot p^j (1-p)^{k-j}. 
\end{equation}
This can be computed efficiently by creating the $m \times m$ matrix $M_{kj} = \binom{k}{j} \cdot p^j (1-p)^{k-j}$ and the vector of probabilities $P'$ is given by the matrix vector product $P' = P \times M$.

\subsubsection{Adding loss}

We take loss into consideration by considering that the auxiliary photons that have to be detected can be lost with a nonzero probability. At each stage, we estimate the optical depth seen by such a photon, $d$, and given a fixed transmission per component $\eta$, the success probability of the computation stage is updated as $p' = p \times \eta^{d \times n}$, where $n$ is the number of photons to be detected at this stage. We recover a scenario with no loss but with smaller success probabilities at each stage.

\section{Optimization procedure for Figure~\ref{fig:results}}
\label{app:Optitim}

The purpose of Figure~\ref{fig:results} is to highlight the tradeoff between the resource efficiency of the schemes and their maximum loss allowed per component to stay in the FTQC regime. More precisely, for a given loss $x$, we want to optimize the resource efficiency of the scheme such that the total transmission of the photons of the resource states is above $92.5\%$ and such that the success probability of having at least one resource state per cycle is $p_{\text{target}} = 99.9\%$. Given the variety of possibilities in terms of resource sharing, doing such optimization can be a tedious task. To simplify, we make the following two assumptions: 
\begin{itemize}
    \item the scheme will be a series of $k$ independent RSG modules, each module having perfect resource sharing, 
    \item each RSG module can have at most $100 000$ sources. This value is arbitrary but gives an upper bound on the maximum amount of sources that can be realistically managed in a perfect resource sharing model.
\end{itemize}

\subsection{Maximum optical depth and maximum module size}

Given a value for the loss $x$, we can deduce from Ineq.~\ref{eq:threshold} the maximum optical depth $D_{\max}$ at our disposal to stay in the FTQC regime: 
\begin{equation}
    D \leq \frac{\log(0.925)}{\log(1-x)}.
\end{equation}
Hence $D_{\max} = \floor*{\frac{\log(0.925)}{\log(1-x)}}$. Then, depending on the scheme, we can derive analytical formulas for $D$ as a function of the size of a module $N_s$. This will give us an upper bound for $N_s$ as well. 

\subsubsection{All-photonic scheme}

We assume for simplicity that $N_s=96m$. Then the number of states at each stage are: 
\begin{itemize}
    \item the number of primates is $K/4$,
    \item the number of GHZ is $K/8$,
    \item the number of building blocks is $K/16$,
    \item the number of half rings is $K/48$,
    \item the number of rings is $K/96$.
\end{itemize}

With a $\ceil*{\log_2(n)}$ sorting network at each stage, the total optical depth is approximately given by 
\begin{equation}
    \begin{split}
        & D = (p_s != 1) \times \log_2(K) + \log_2(K/4) \\ 
        & + \log_2(K/8) + \log_2(K/16) + \log_2(K/48) + 3
    \end{split}
\end{equation} 
where the boolean $(p_s != 1)$ accounts for the fact that there is no need to multiplex the photons if the source is deterministic.

From this we can deduce the maximum number of sources allowed per module: 
\begin{equation}
    \begin{split}
        (N_s)_{\max} = \min \Big( & \floor*{2^{D_{\max}-3 + \log_2(24576)/(4+(p_s != 1))}}, \\
        & 100000 \Big) .
    \end{split}
\end{equation}

\subsubsection{Caterpillar source}

A similar procedure applies with the caterpillar source. The total optical depth $D$ does not change, what changes is the depth of the scheme: 
\begin{itemize}
    \item we still account for the multiplexing of the final resource states in constant depth $3$, 
    \item and all that remains is multiplexing the caterpillar states after the first round of fusions. If we have $N_s$ sources in the module, then we need $\ceil*{\log_2(N_s)}$ depth to do the multiplexing. 
\end{itemize}

Overall, the maximum number of sources allowed per module is 

\begin{equation}
    (N_s)_{\max} = \min\left(2^{D_{\max}-3}, 100000 \right).
\end{equation}

\subsection{Number of modules and optimization problem}

Once we fix the number of sources $N_s$ in one module, we can compute the success probability of the module $p_{\text{rsg}}$ using the process described in Appendix~\ref{sec:simulation_perfect_sharing}. Then, to reach the target probability $p_{\text{target}} = 0.999$, we need to repeat the module $k$ times, with $k$ such that $1 - (1-p_{\text{module}})^k \geq p_{\text{target}}$, i.e. 
\begin{equation}
    k = \ceil*{\frac{\log(1-p_{\text{target}})}{\log(1-p_{\text{module}})}}. 
\end{equation}  

Then the resource efficiency is a function of $N_s$ and $k$.

\subsubsection{All-photonic scheme}

Writing $N_s = 96m$, the total number of sources required is $k \times m \times (96 + 18 \times 4) = 168km$, where we account for $72m$ extra sources per module to add the photons that boost the fusions ($72$ extra photons for a maximum of $m$ final resource states). This implies
\begin{equation}
    \eta_r = \frac{24}{168 \times k \times m}.
\end{equation}

\subsubsection{Caterpillar source}

With $N_s= 3m$ sources in our module, the total number of sources is given by $k \times m \times 4$, where $m$ additional sources per module accounts for the production of Bell pairs to boost the fusions (one source per resource state). The total number of photons consumed is given by $k \times m \times 69$, where $51$ photons are produced per set of $3$ sources in one module and $18$ additional photons to boost the $9$ fusions of the scheme are produced by the fourth source. This implies 
\begin{equation}
    \eta_r = \frac{24}{69 \times k \times m}.
\end{equation} 

\subsubsection{Optimization problem}

In both cases, we want to minimize $km$ (the resource efficiency) with the constraints that $ 1 - (1-p_{\text{rsg}})^k \geq p_{\text{target}}$ (the success probability of the full scheme is $0.999$) and $\alpha \cdot m \leq (N_s)_{\max}$ (the optical depth $D$ is such that we are in the FTQC regime) where $\alpha = 96$ for the all-photonic scheme and $\alpha=3$ with the caterpillar source. 

\section{RUS module Monte Carlo simulation details}
\label{app:MonteCarlo}

Since the RUS module architecture we propose has many fewer sources compared to the all-photonic architecture, it is convenient to simulate and optimize it using a Monte Carlo simulation. The aim of this simulation is to predict the minimum number of internal clock cycles and number of sources needed to ensure that the RSG success probability reaches $p_\mathrm{rsg}=0.999$.

We assume that the time required to attempt a single CZ gate sets the internal clock cycle, as this operation demands hardware-level feedforward. We neglect the time required for single-qubit operations, as their contribution is negligible compared to the duration of two-qubit interactions. Deterministic single-qubit gates (e.g., Hadamards for LC) can be performed between RUS gates without extending the internal clock cycle, which is set by the feedforward latency required to halt photon emission after a successful RUS CZ gate. These single-qubit gates can already be implemented in under 1~ns~\cite{huet2024deterministic}, faster than even optimistic feedforward estimates limited by classical electronics. Pauli or Clifford corrections following each RUS CZ attempt can be commuted to the end of the graph construction or perhaps absorbed into the Pauli frame during fusion network execution. Unlike RUS operations, single-qubit measurements do not need to be halted in real time, and repeated cycling enables high-fidelity readout in well under one clock cycle. Finally, in our simulation, we allow for two full internal clock cycles to perform spin-to-photon mapping, which is ample time for photon emission and spin measurement/reset.

We perform the simulation in four stages. The first stage builds 4 subgraphs of three spin qubits each using two CZ gates for each subgraph applied in two consecutive rounds of RUS. The second stage takes two complete subgraphs of the first stage and attempts to connect them using two additional CZ gates that are applied during a single round of RUS. The third stage takes two complete subgraphs from the second stage and attempts to connect them using two CZ gates applied during a single round of RUS. For convenience, we add a fourth deterministic stage to capture the cost in time to emit the entangled state from the sources. It begins once the final edge is established and takes two internal clock cycles to complete.

For each round of RUS, we iterate through internal clock cycles and pass any complete subgraphs to the next stage. To sample each attempt to establish an edge, we first determine if both photons were not lost by sampling a random number between 0 and 1 and comparing it to $\eta_\text{rus}^2$, where $\eta_\text{rus}$ is the end-to-end transmission efficiency of the RUS gate. If at least one photon was lost, an erasure error occurred, and the stage is reset conditioned on logic to determine the recovered state of previous stages (see Paragraph ``Erasure logic'' below for more details). Otherwise, we sample with a 50\% probability that the gate was successful.

Every $\tau$ internal clock cycles, we check to see if at least one full graph is complete. If so, the cycle is marked as a success and the 12 sources are reset to the first stage to begin a new graph. If no graph is complete, the cycle is marked as a failure but no graphs are reset. This latter behavior captures the temporal multiplexing property of the RUS scheme.

We repeat the above process to acquire samples of $10^5$ RSG cycles and estimate $p_\mathrm{rsg}$. For a given value of $\beta$ and $x$, we then determine the value of $\tau$ and the number of sources such that 99.9\% of all RSG cycles were successful, ie.~$p_\mathrm{rsg}=0.999$. The value $\tau=r_0/r$ along with the required number of sources $N_0$ is then used to compute the resource efficiency using $\eta_R = 24/(N_0 \tau)$.

In the case where we use a single group of 12 sources and no spatial multiplexing as in Figure 4a of the main text (solid curve), we take $\eta_\mathrm{rus}=\eta_T=\beta(1-x)$, where $x$ is the loss from a single $1\times (d+1)$ active switch (GMZI) needed to route the photon to at most degree $d=4$ other sources plus the output mode.

To simulate the spatial multiplexing used in Figure 4a of the main text (dashed curve), we run multiple independent 12-source constructions in parallel. However, at the end of the RSG cycle, we select and reset at most one completed graph, storing any remaining completed graphs for the next cycle. For this simulation, we use $\eta_\mathrm{rus}=\beta(1-x)$ and $\eta_T=\beta(1-x)^2$ where a second $n\times 1$ switch is needed to recombine the $n$ multiplexed groups into a single output mode. Note that delay lines are not necessary since the state can be emitted only after the switches are reconfigured. Thus, in principle, all switching, routing, and entangling could be accomplished using a single photonic integrated circuit.

To simulate the spatial resource sharing used in Figure 4b of the main text, we follow the above procedure but assume that the next stage can begin as soon as any two subgraphs of the current stage are complete. The cost of this is to increase the degree of connectivity $d$ to route photons between sources. As a consequence, it may become favorable to go off chip to avoid exponentially many mode crossings with increasing connectivity and so we assume $\eta_\mathrm{rus}=\beta(1-x)^3$ to account for extra fibre couplings.

\begin{figure}[t]
    \centering
    \includegraphics[width=1\linewidth, trim= 350 350 230 100, clip]{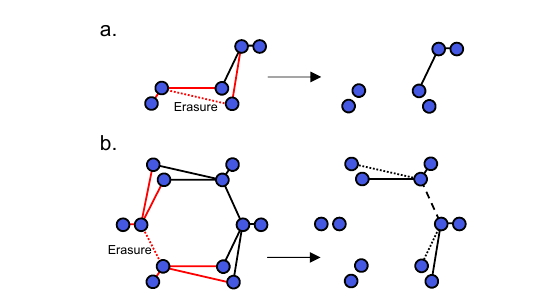}
    \caption{\textbf{a.} Recovered state after a single erasure error during stage 2. \textbf{b.} Recovered state after a single erasure error during stage 3. In both panels, the dashed red edge illustrates the edge where a photon was lost and the red solid edges show the destroyed edges due to erasure errors. To understand the effect of erasures, note that a photon loss during a RUS gate leads to a dephasing of the quantum emitters' qubits (see~\cite{hilaire2024enhanced} for the derivation). To obtain a graph state again, one then performs Z measurements on the affected qubits, followed by a Hadamard gate (see e.g. \cite{hein2006entanglement} for a comprehensive walkthrough of how Pauli measurements affect graph states). In panel b the black dotted and dashed edges are not destroyed and can only improve the success probability compared to the recovered two stage 1 subgraphs: the dotted edges contributes in stage 2 while the dashed edge contributes in stage 3.}
    \label{fig:erasure}
\end{figure}

\paragraph*{Erasure logic.}

Since the sources act as a quantum memory, and erasures do not always completely destroy the graph, it is possible to recover partially completed stages after some erasures. To demonstrate this, we consider two such cases to help alleviate the impact of photon loss (see Fig.~\ref{fig:erasure}).

If a single erasure occurs during stage 2 while the other edge succeeds, the resulting graph will contain a completed subgraph from stage 1. Thus, retrying stage 2 requires just one stage 1 subgraph to be built. However, if both RUS gates fail during the same round of RUS, both subgraphs will be completely destroyed.

If one erasure occurs during stage 3, half of the completed graph will be destroyed. Although this remaining graph does not correspond to any subgraph of a previous stage, it has all the required edges of two completed subgraphs from stage 1 plus additional edges. These additional edges may be destroyed if more erasures occur, or if they survive until stage 2 or 3, they will only benefit the construction. Hence, a pessimistic assumption is that we can recover two completed subgraphs in stage 1.
 
Note that there are further ways to exploit state recovery and, unlike equivalent all-photonic strategies, does not increase the hardware complexity or optical depth.

\section{RSG probability transition}
\label{app:prsg}

In the main text, we fix $p_\mathrm{rsg}=0.999$ to ensure we probe a regime where independent loss dominates. To verify that this assumption does not bias the conclusions of our study, it is useful to establish a relationship between $\eta_R$ and $p_\mathrm{rsg}$, which allows us to bound improvements on $\eta_R$ that would result from decreasing $p_\mathrm{rsg}$.

Consider, for example, a device that produces a resource state with $p_\mathrm{rsg}=0.97$ and a resource efficiency of $\eta_R=0.02$. Using a simple spatial multiplexing strategy---run two such devices in parallel and combine the outputs with a single switch---we obtain a new RSG with $p_\mathrm{rsg}=1 - (1 - 0.97)^2=0.9991$ and $\eta_R=0.01$. This demonstrates that when $p_\mathrm{rsg}$ is close to 1, additional resources can exponentially suppress the failure probability of the RSG. Importantly, the vast majority of the resources are required simply to bring $p_\mathrm{rsg}$ above 0.5. The remaining improvement from $p_\mathrm{rsg}=0.5\rightarrow 0.999$ requires at most one order of magnitude increase in resources. Conversely, decreasing $p_\mathrm{rsg}$ from 0.999 to 0.9 will increase $\eta_R$ by at most a factor of 3.

In practice, advanced multiplexing strategies such as resource sharing can achieve the same exponential suppression of $1-p_\mathrm{rsg}$ at a resource cost significantly lower than the naive case described above. For highly optimized architectures, this sharp transition is even more pronounced, such that resource cost difference between $p_\mathrm{rsg}=0.9$ and $p_\mathrm{rsg}=0.999$ is almost negligible.

For example, in our simulation of the RUS scheme with resource sharing, spatial and temporal multiplexing, $\beta = 95\%$, $x = 0.1\%$, and $r / r_0 = 1$ (1~GHz maximum operation), we find that a resource efficiency of $\eta_R = 10.8\%$ (222 sources) is needed to achieve $p_\mathrm{rsg} = 0.9$, while achieving $p_\mathrm{rsg} = 0.999$ requires only $\eta_R = 8.3\%$ (288 sources). Thus, relaxing $p_\mathrm{rsg}$ from 0.999 to 0.9 improves $\eta_R$ by a factor of 1.3, far below the upper bound of 3. This reflects the fact that $\eta_R$ is already close to $\tilde{\eta}_R$ due to the efficient multiplexing strategy.

For the all-photonic scheme simulation, we assume perfect resource sharing for all stages except the final stage, which uses simple spatial multiplexing for the final boost to the target value of $p_\mathrm{rsg}$. As a result, reducing $p_\mathrm{rsg}$ from 0.999 to 0.9 increases $\eta_R$ by the full factor of 3. This highlights the difficulty of achieving $\eta_R \approx \tilde{\eta}_R$ in the all-photonic approach, due to the large number of resources that must be shared efficiently. Conversely, relaxing $p_\mathrm{rsg}$ provides proportionally greater improvements in this case.

\section{Two-qubit operations}
\label{app:twoqubitinteractions}

To provide a rough comparison of the expected error sensitivity between the proposed architectures, we estimate the number of effective two-qubit interactions required to generate a single (2,2) Shor-encoded 6-ring resource state. We focus only on the number of two-qubit interactions as an indicator of error susceptibility, as these operations tend to be significantly more error-prone than single-qubit gates and measurements in photonics. Moreover, this provides a simple and interpretable metric to compare spin-based and all-photonic schemes on equal footing. We furthermore distinguish between two-photon interactions (e.g., fusion gates requiring high photon indistinguishability) and spin-photon interactions (e.g., entangled photon emissions) to capture the trade-off between dependence on photonic quality and spin quality. Table.~\ref{tab:num2qubit} summarizes this analysis. In the following, we detail the assumptions and calculations behind these values.

\begin{table}[h!]
    \centering
    \begin{tabular}{c||c|c|c}
         Scheme & Two-photon & Spin-photon &  Combined \\\hline\hline
         All-photonic & 102 & 0 & 102 \\
         Hybrid & 54 & 48 & 102\\
         Caterpillar & 18 & 60 & 78 \\
         RUS Module & 28 & 80 & 108 \\
    \end{tabular}
    \caption{Estimates of the number of effective two-qubit interactions required to produce a Shor-encoded (2,2) 6-ring resource state.}
    \label{tab:num2qubit}
\end{table}

While the resource analysis in the main text considers the total average number of parallel attempts needed to generate a resource state, the error affecting a successfully constructed state depends only on the operations that actually contributed to its creation. For the all-photonic, hybrid, and caterpillar schemes, this number is fixed, as failed attempts are discarded and do not affect fidelity. In the RUS Module scheme, failed gate attempts can be retried, so we report the average number of operations contributing to each successful gate.

For the all-photonic scheme, each resource state requires 12 successful 4-GHZ seed states, 12 type-I fusion gates, and 6 type-II fusion gates. Since we do not explicitly model the error mechanisms involved for these operations, we estimate their contribution to the overall error by counting effective two-photon interactions using the number of interfering photons. This estimate is based on a model of photon distinguishability, where the success probability of a multi-photon interference process involving $n$ photons scales as $V^{n/2}$ where $V$ is the two-photon interference visibility \cite{annoni2025incoherent}. Hence, for $n=8$ photons interfering to produce a single 4-GHZ, we take $n/2=4$ two-photon interactions. Each boosted fusion gate requires the interference of two graph photons and four auxiliary photons, thus we approximate it as three effective two-photon interactions. This leads to a total of
$12\times4+18\times3=102$ effective two-photon interactions.

In the hybrid scheme, 4-GHZ seeds are generated using spin-photon entanglement rather than multi-photon interference. Each emission of a photon entangled with the spin constitutes an effective spin-photon entangling gate---equivalent to a CNOT operation in the sequential generation picture~\cite{schon2007sequential}. Since each 4-GHZ seed requires the emission of 4 entangled photons, the total number of spin-photon interactions required is $12\times 4 = 48$. The rest of the procedure mirrors the all-photonic scheme, involving 12 type-I and 6 type-II fusion gates, for a total of $18\times3=54$ two-photon interactions. Thus, this hybrid scheme has the same total number of 102 two-qubit interactions as the all-photonic scheme, but with the resources nearly equally split between photon-photon and spin-photon interactions.

The caterpillar scheme is even more spin-focused. Each of the three seed states requires 14 spin-photon operations. These seed states are then connected using 9 boosted type-II fusion gates. Each boosted fusion interferes the two photonic qubits from the seed states using an auxiliary Bell pair, which is itself generated using two additional spin-photon emissions from a separate caterpillar source. In total, this gives $3\times 14 + 9\times 2 =60$ effective spin-photon interactions and $2\times 9=18$ photon-photon interactions. The overall interaction count is lower than in the previous schemes, with just 78 total two-qubit operations, and the majority of these involve the spin.

For the RUS Module scheme, each resource state requires 14 successful RUS gates and a final 24 photon emission events. Each RUS attempt requires two spin-photon interactions and one photon-photon interaction. On average, two attempts are necessary to successfully establish an edge. Thus, the RUS Module scheme requires $2\times2\times14+24=80$ spin-photon interactions and $2\times 14=28$ two-photon interactions, on average.

These estimates highlight a trade-off between photonic and spin-based operations in the generation of encoded resource states. In particular, the observed gains in resource efficiency correspond to an increased reliance on spin-photon interactions, with the RUS Module scheme placing the greatest demands on the spin quality. Conversely, these same schemes reduce the reliance on high-fidelity photon-photon interference, such as indistinguishability and multiphoton interference visibility.

\bibliography{bib}

\end{document}